\documentclass[referee]{raa}            

\usepackage{graphicx,times}             

\begin{document}

   \title{Comparisons of Jet Properties between GeV Radio Galaxies and Blazars
}

   \volnopage{Vol.0 (200x) No.0, 000--000}      
   \setcounter{page}{1}          

   \author{Zi-Wei Xue
      \inst{1,2}
   \and Jin Zhang
      \inst{1,3,4}
   \and Wei Cui
      \inst{3}
   \and En-Wei Liang
      \inst{4}
   \and Shuang-Nan Zhang
      \inst{1,5,2}
   }

   \institute{Key Laboratory of Space Astronomy and Technology, National Astronomical Observatories, Chinese Academy of Sciences,
             Beijing 100012, China; {\it jinzhang@bao.ac.cn}\\
        \and
          University of Chinese Academy of Sciences, Beijing 100049, China   \\
        \and
        Department of Physics and Astronomy, Purdue University, West Lafayette, IN 47907, USA  \\
             \and
            Guangxi Key Laboratory for Relativistic Astrophysics, Department of Physics, Guangxi University, Nanning 530004, China\\
             \and
            Key Laboratory of Particle Astrophysics, Institute of High Energy Physics, Chinese Academy of Sciences, Beijing 100049, China\\
   }

   \date{Received~~201X month day; accepted~~201X~~month day}

\abstract{ We compile a sample of spectral energy distribution (SED) of 12 GeV radio galaxies (RGs), including eight FR I RGs and four FR II RGs. These SEDs can be represented with the one-zone leptonic model. No significant unification as expected in the unification model is found for the derived jet parameters between FR I RGs and BL Lacertae objects (BL Lacs) and between FR II RGs and flat spectrum radio quasars (FSRQs). However, on average FR I RGs have the larger $\gamma_{\rm b}$ (break Lorentz factor of electrons) and lower $B$ (magnetic field strength) than FR II RGs, analogous to the differences between BL Lacs and FSRQs. The derived Doppler factors ($\delta$) of RGs are on average smaller than that of balzars, which is consistent with the unification model that RGs are the misaligned parent populations of blazars with smaller $\delta$. On the basis of jet parameters from SED fits, we calculate their jet powers and the powers carried by each component, and compare their jet compositions and radiation efficiencies with blazars. Most of the RG jets may be dominated by particles, like BL Lacs, not FSRQs. However, the jets of RGs with higher radiation efficiencies tend to have higher jet magnetization. A strong anticorrelation between synchrotron peak frequency  and jet power is observed for the GeV RGs and blazars in both the observer and co-moving frames, indicating that the ``sequence" behavior among blazars, together with the GeV RGs, may be dominated by the jet power intrinsically.
\keywords{galaxies: active---galaxies: general---galaxies: jets---gamma rays: galaxies---radiation mechanisms: non-thermal}
}

   \authorrunning{Z.-W. Xue et al. }            
   \titlerunning{Comparisons of Jet Properties between GeV Radio Galaxies and Blazars }  

   \maketitle
%
%
\section{Introduction}           
\label{sect:intro}

Radio galaxies (RGs) belong to a sub-class of active galactic nuclei (AGNs). It was 40 years ago that Fanaroff \& Riley (1974) classified RGs into two groups: Fanaroff \& Riley Class I (FR I) and Class II (FR II) RGs according to their radio morphology; FR I RGs are core-dominated with ``edge-dimmed" radio lobes and FR II RGs are lobe-dominated with ``edge-brightened" radio lobes. The classification of radio morphology for RGs is consistent with the radio power distinction: RGs with radio powers lower than $10^{32}$ erg s$^{-1}$ Hz$^{-1}$ at 408 MHz exhibit almost exclusively FR I morphologies while RGs with radio powers higher than $10^{34}$ erg s$^{-1}$ Hz$^{-1}$ at 408 MHz show almost exclusively FR II morphologies (Zirbel \& Baum 1995). However, in both classifications of radio power and radio morphology, there is a considerable overlap over which RGs can be identified as either an FR I or an FR II RG (Baum et al. 1988; Owen \& Laing 1989; Morganti et al. 1993; Zirbel \& Baum 1995). Besides the differences in radio power and radio morphology, the strong dichotomy of RGs in optical properties has also been studied by many authors (e.g., Zirbel \& Baum 1995; Buttiglione et al. 2009; Baldi \& Capetti 2009): strong emission lines occur in the more powerful FR II RGs, but the weaker FR I RGs tend to have no emission line. FR I and FR II RGs may also correspond to the low-excitation RGs (LERGs) and high-excitation RGs (HERGs), respectively, although the low/high-excitation RGs do not have one-to-one correspondence with the FR I/FR II categories (Hine \& Longair 1979; Laing et al. 1994; Hardcastle et al. 2009). FR I and FR II RGs may intrinsically have different accretion modes (e.g., Wu et al. 2008; Xu et al. 2009), which may also be unified with BL Lacertae objects (BL Lacs) and flat spectrum radio quasars (FSRQs, Xu et al. 2009; Zhang et al. 2014, 2015). The physical reasons for the RG division are still unclear and are also highly debated. However, fuelling mechanism and merging history may play important roles (Hardcastle et al. 2007; Saripalli 2012).

So far, only four FR I RGs have been detected at very high energy (VHE) $\gamma$-ray band (TeV band), i.e., M87 (Aharonian et al. 2003), Cen A (Aharonian et al. 2009), IC 310 (Aleksi\'{c} et al. 2010), NGC 1275 (Aleksi\'{c} et al. 2012). The first confirmed GeV RG is Cen A, which is the only GeV source not belonging to the blazar class in the third EGRET Catalog of $\gamma$-ray sources (Hartman et al. 1999). Now there are 14 RGs detected at GeV band with \emph{Fermi}/LAT (Ackermann et al. 2015). The $\gamma$-ray emission has been detected and confirmed in the radio lobes of Cen A (Abdo et al. 2010c), which is the first detection of $\gamma$-ray emission in the large-scale extended regions of AGNs. It also confirms that there are detectable $\gamma$-ray emission for \emph{Fermi}/LAT in the large-scale jets of AGNs (Zhang et al. 2009, 2010). Hence the investigation of radiation mechanisms and the locations of $\gamma$-ray emission for GeV RGs is very important.

It is well known that most of confirmed GeV AGNs are blazars (Ackermann et al. 2015), which are divided into BL Lacs and FSRQs according to the strength of optical emission lines. Their spectral energy distributions (SEDs) are dominated by their jet emission and can be explained with the one-zone leptonic models (e.g., Maraschi et al. 1992; Sikora et al. 1994; Ghisellini et al. 1996; Ghisellini et al. 2009; Sikora et al. 2009; Zhang et al. 2012, 2014, 2015; Chen et al. 2012; Liao et al. 2014). The observed SEDs of RGs resemble that of blazars; they show bimodal feature and can be also explained well by the one-zone leptonic models (Abdo et al. 2009a, 2009b; Aleksi\'{c} et al. 2014; Fukazawa et al. 2015). According to the unification models for radio loud (RL) AGNs, BL Lacs are associated with FR I RGs, whereas FSRQs are usually linked with FR II RGs (Urry \& Padovani 1995), i.e., the RGs are the parent populations of blazars with large viewing angles and small Doppler factors ($\delta$). Based on the large sample, we have investigated the jet properties of GeV blazars in our previous works (Zhang et al. 2012, 2014, 2015). Study on the jet properties in different Doppler amplification systems and comparisons of jet properties between blazars and GeV RGs in both the observer and co-moving frames are important to understand the jet physics and the unification models.

In this paper, we compile a sample of SEDs for GeV RGs to study the radiation mechanisms and physical properties of their jets, and explore the unification model of RL AGNs by comparing the jet properties with that of a balzar sample. The sample and the observed SEDs of GeV RGs are presented in Section 2.  The model and SED fitting are described in Section 3. Comparisons of jet properties between GeV RGs and blazars are presented in Section 4. A summary is given in Section 5.

\section{Sample and Data}

\begin{figure}
\includegraphics[width=37mm,height=34mm]{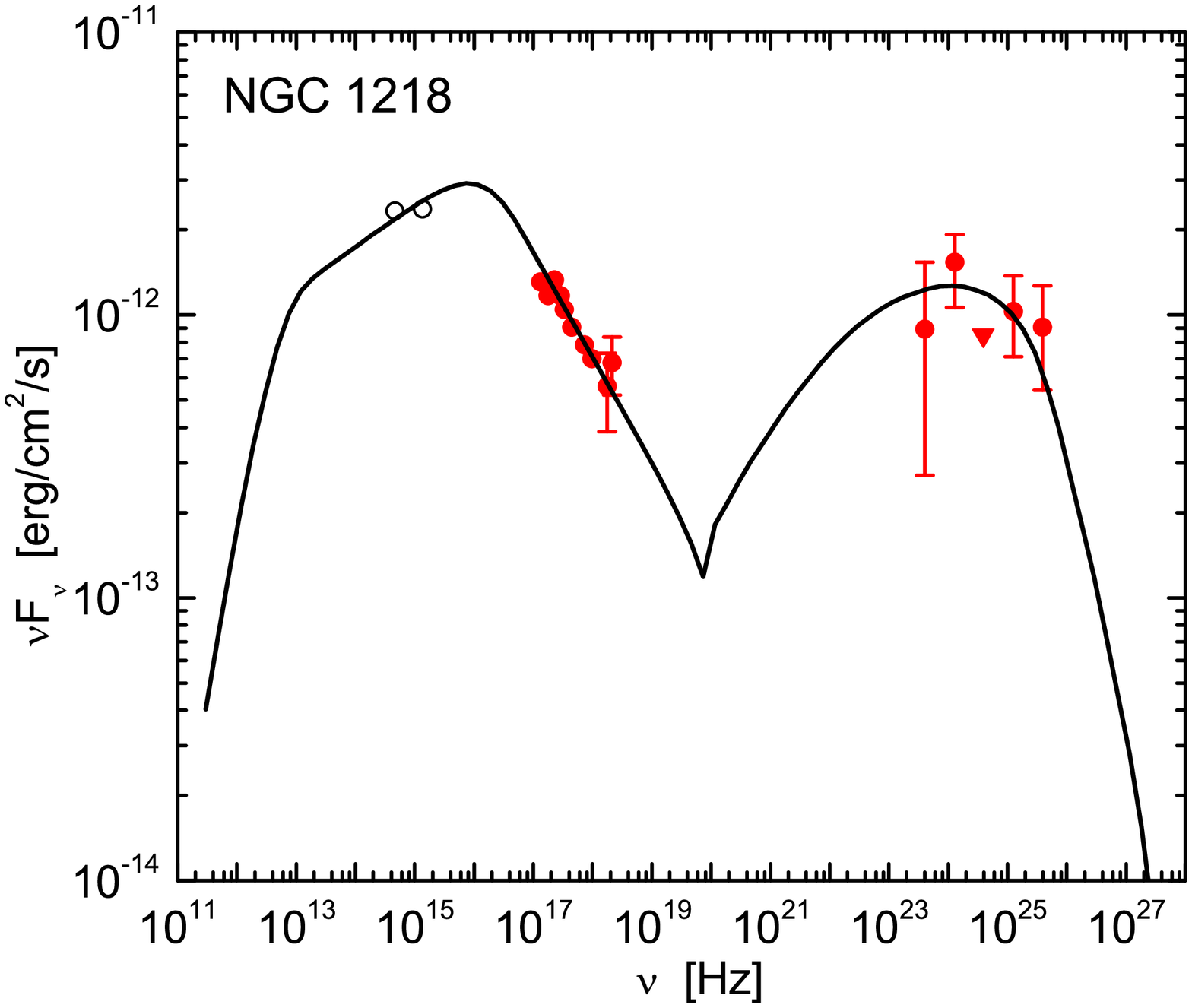}
\includegraphics[width=37mm,height=34mm]{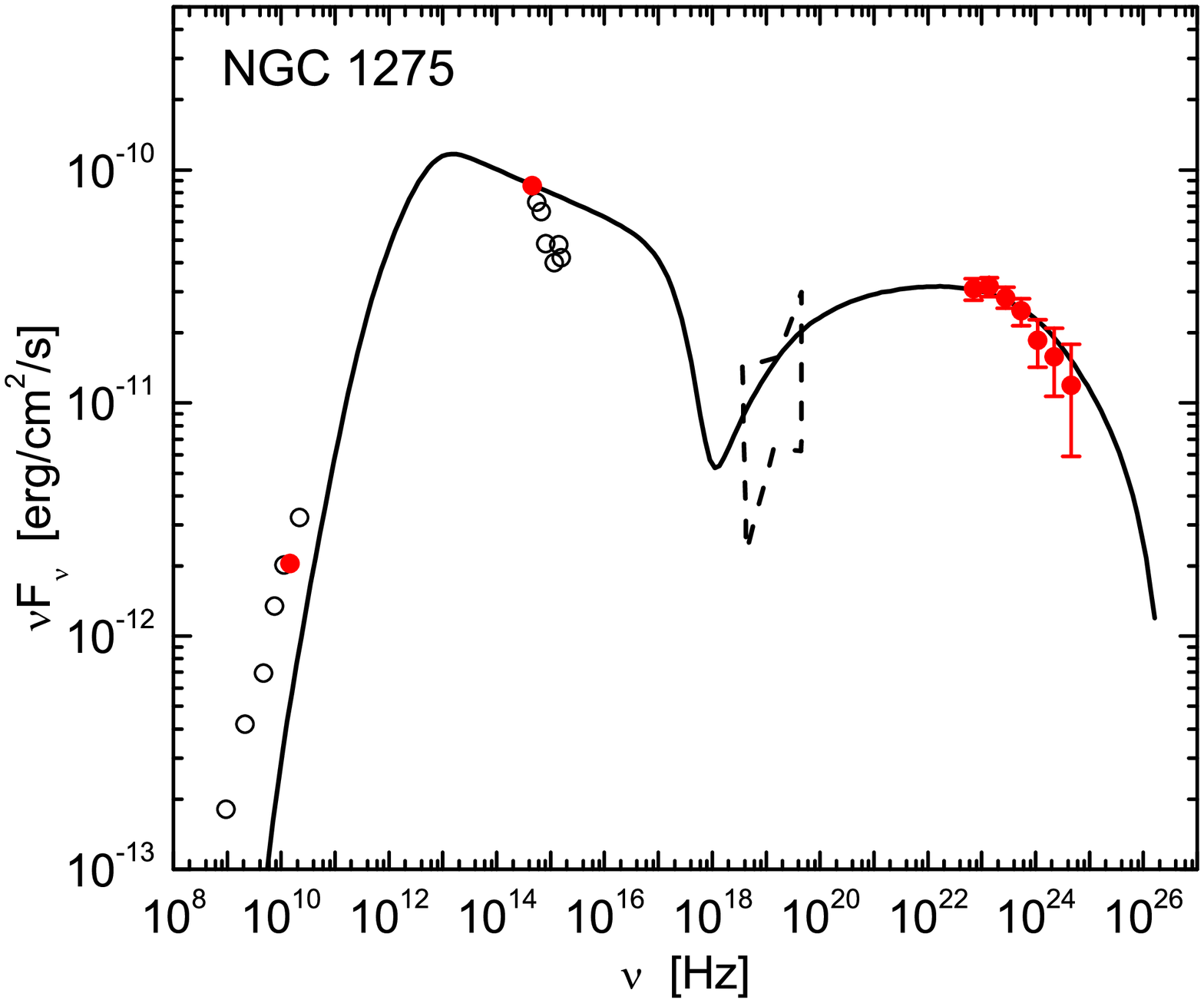}
\includegraphics[width=37mm,height=34mm]{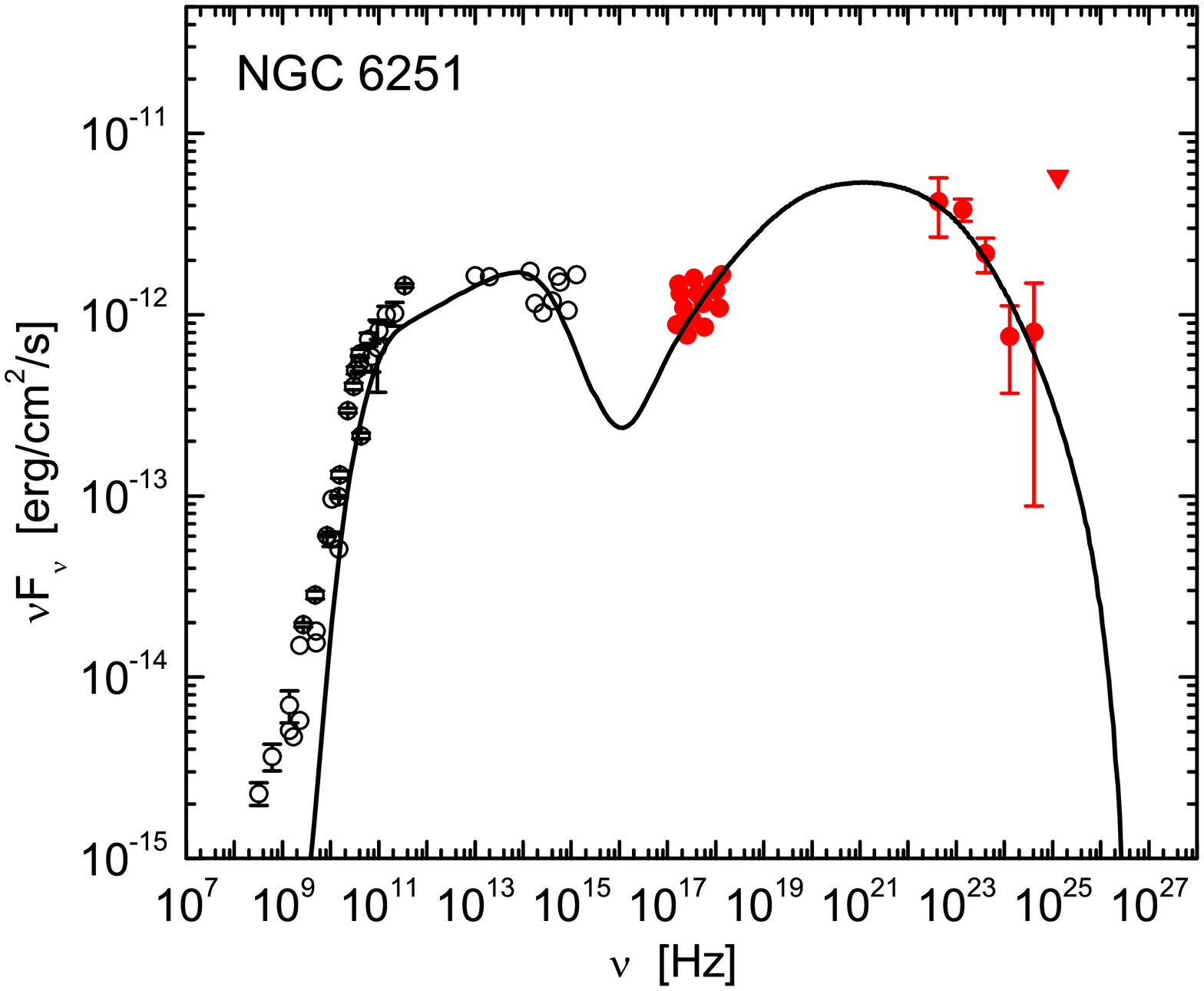}
\includegraphics[width=37mm,height=34mm]{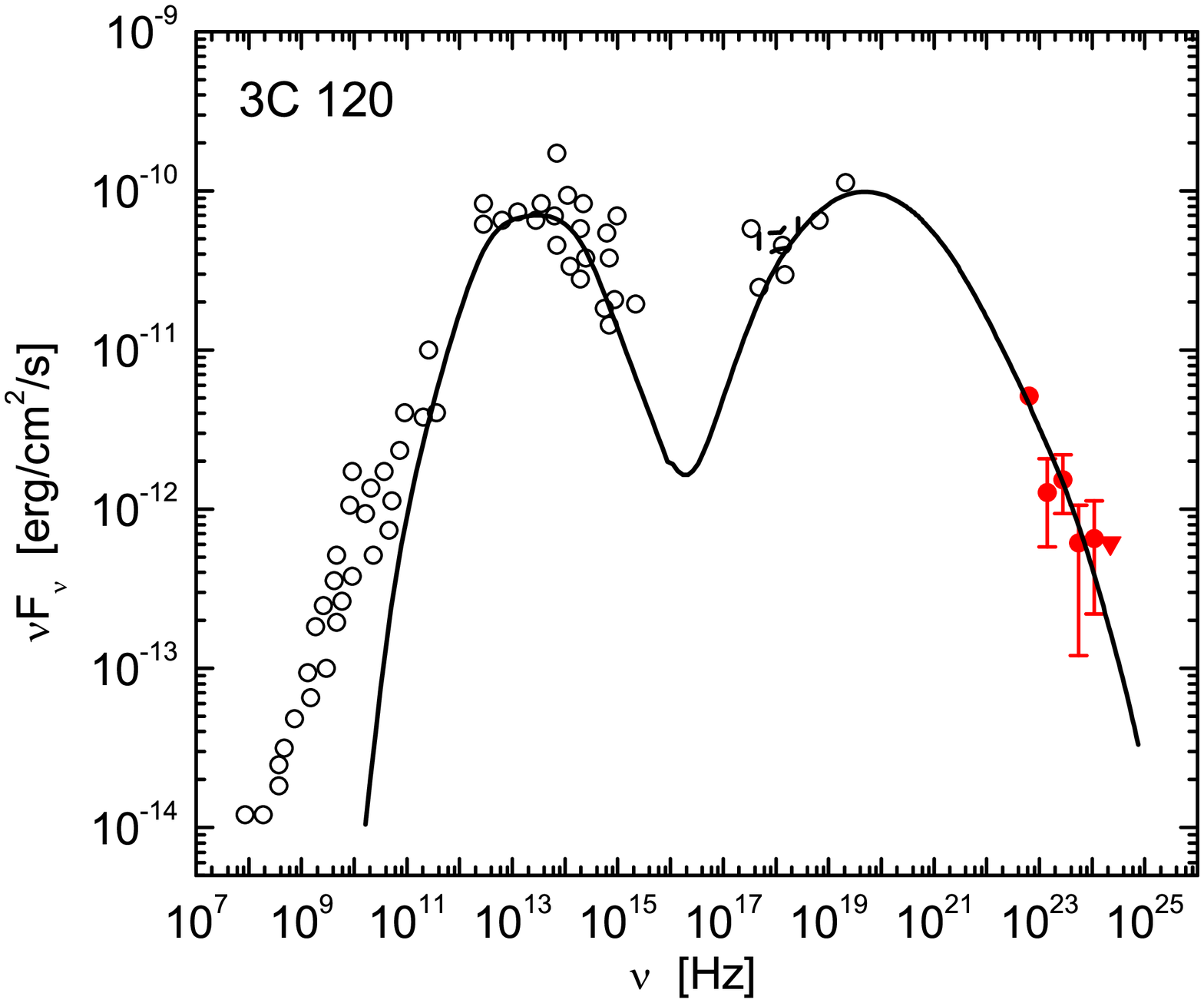}\\
\includegraphics[width=37mm,height=34mm]{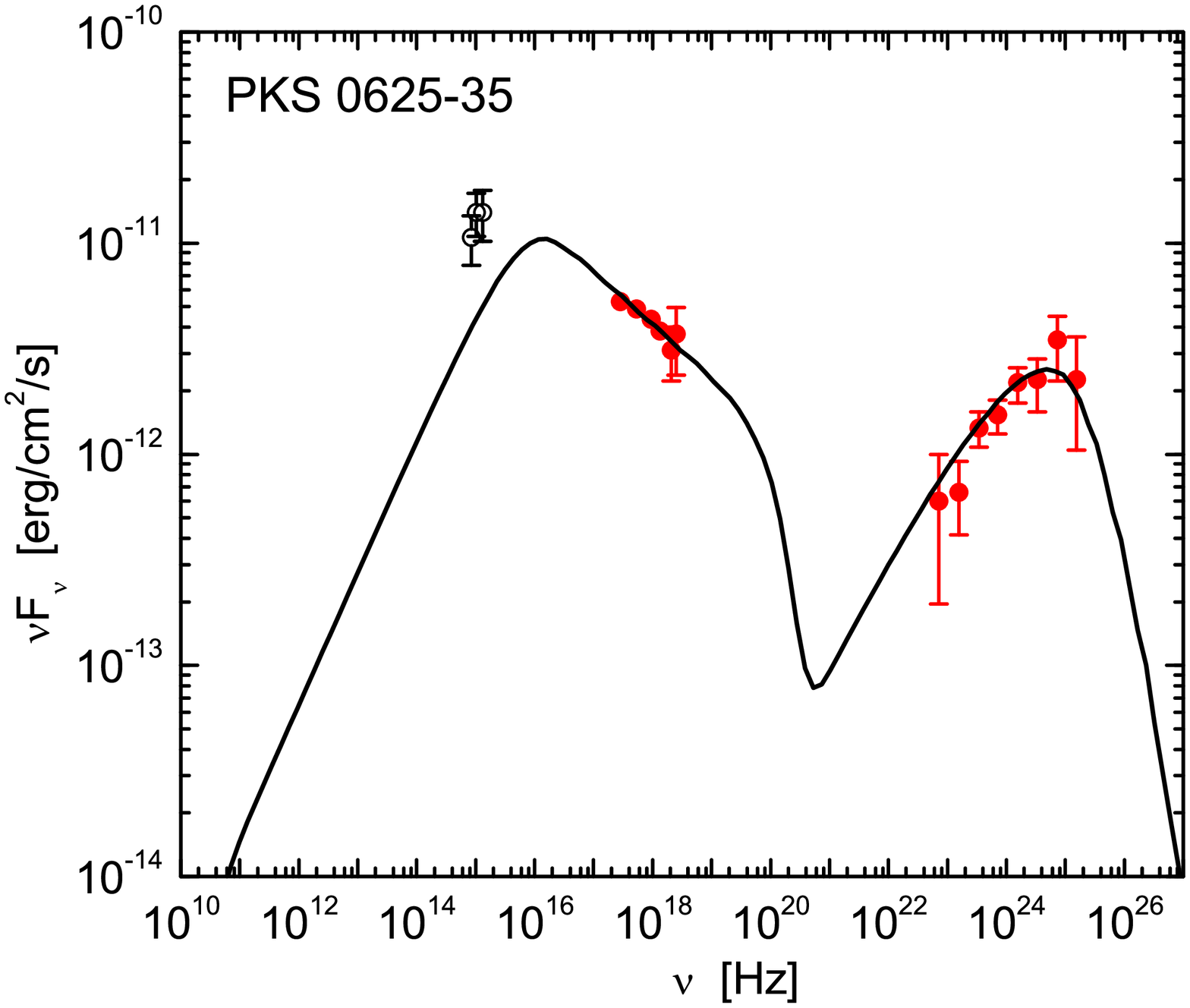}
\includegraphics[width=37mm,height=34mm]{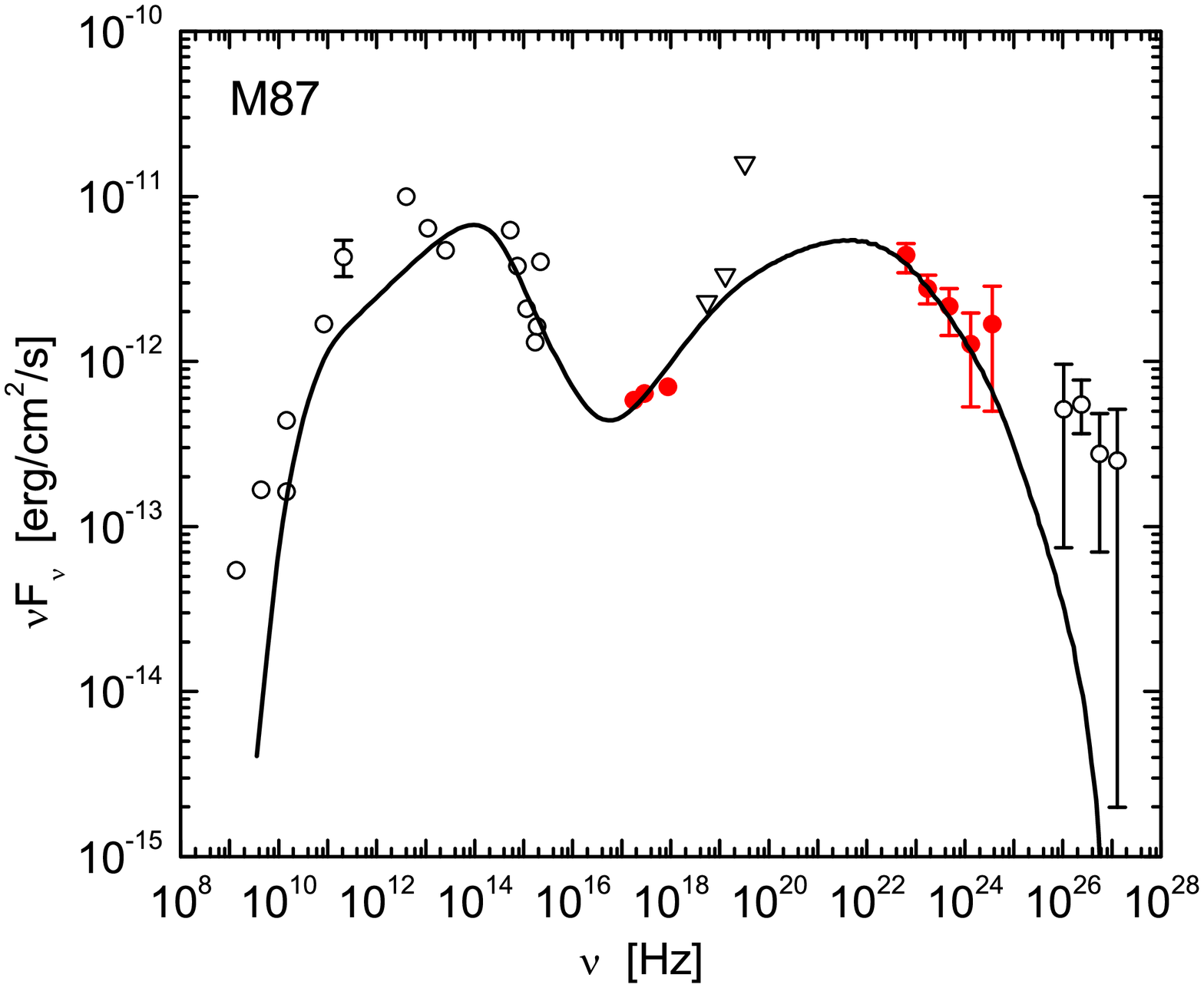}
\includegraphics[width=37mm,height=34mm]{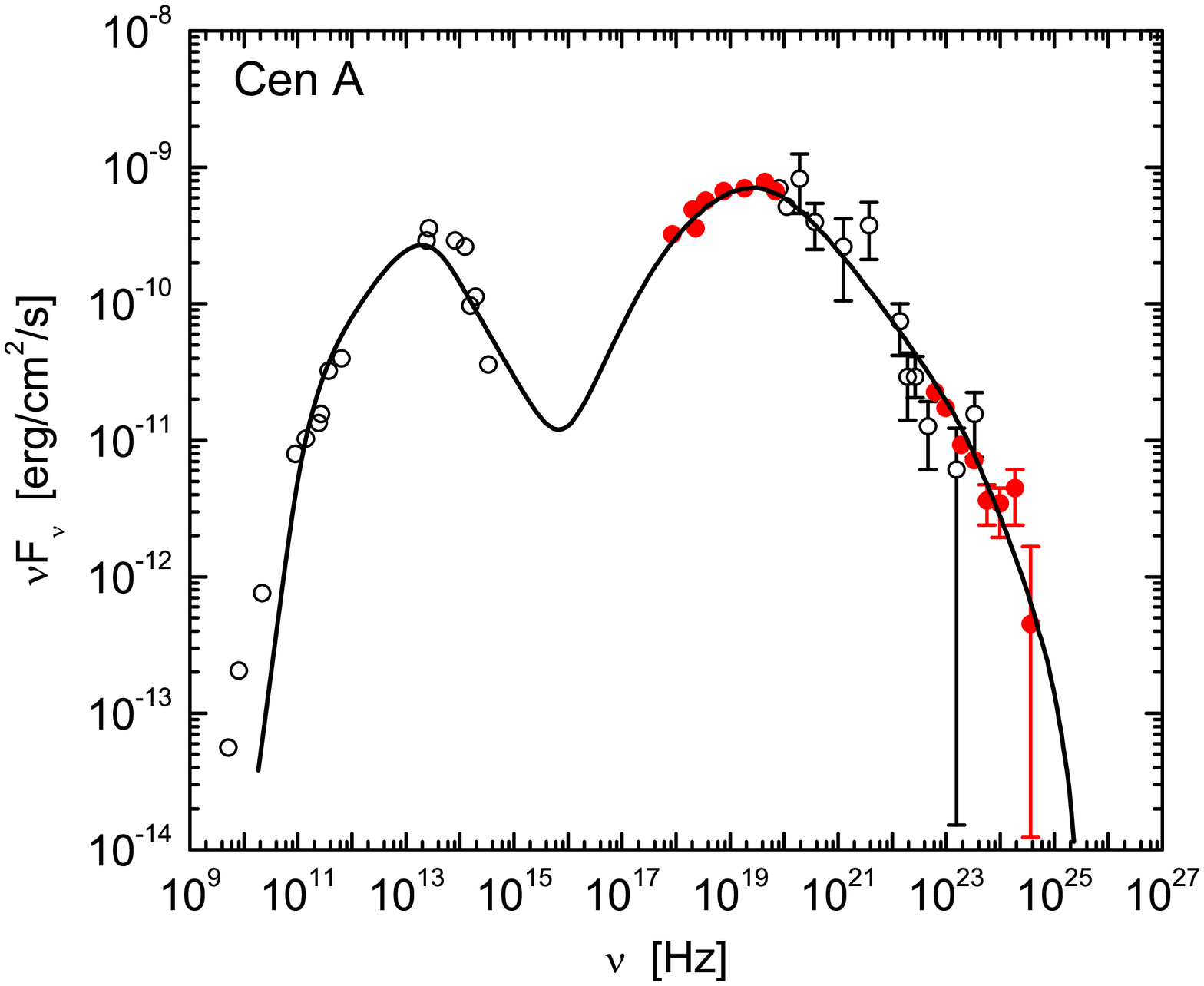}
\includegraphics[width=37mm,height=34mm]{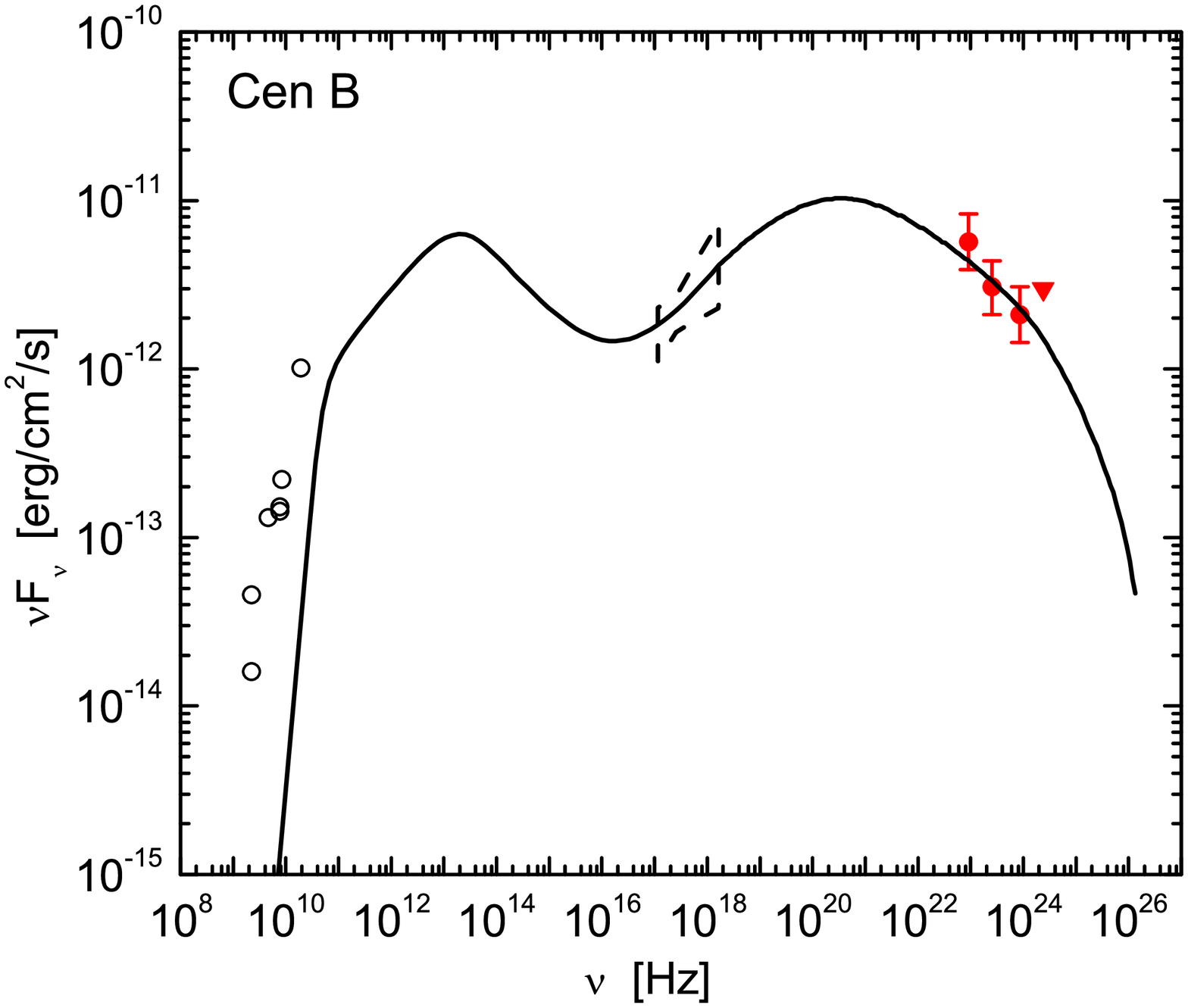}\\
\includegraphics[width=37mm,height=34mm]{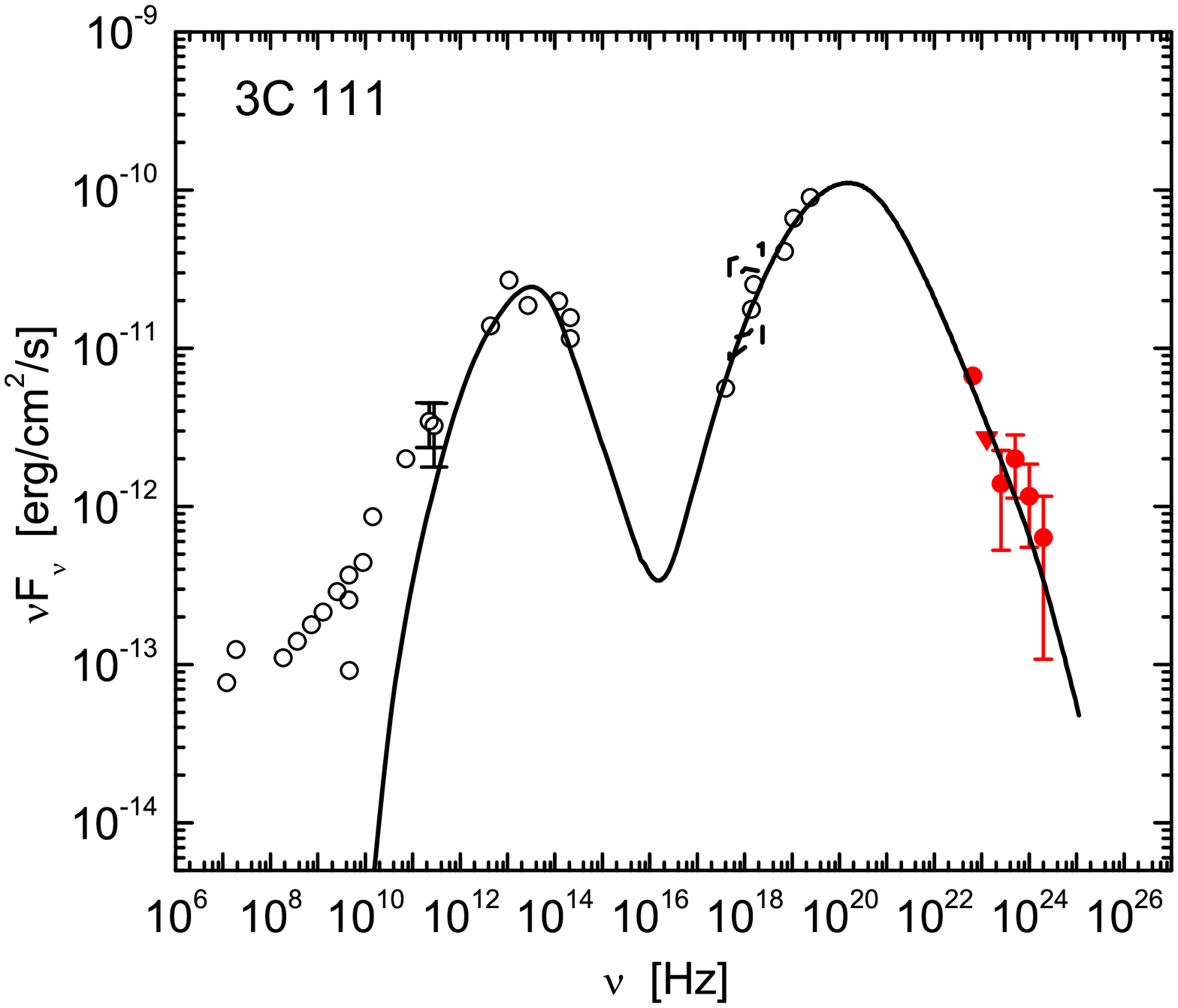}
\includegraphics[width=37mm,height=34mm]{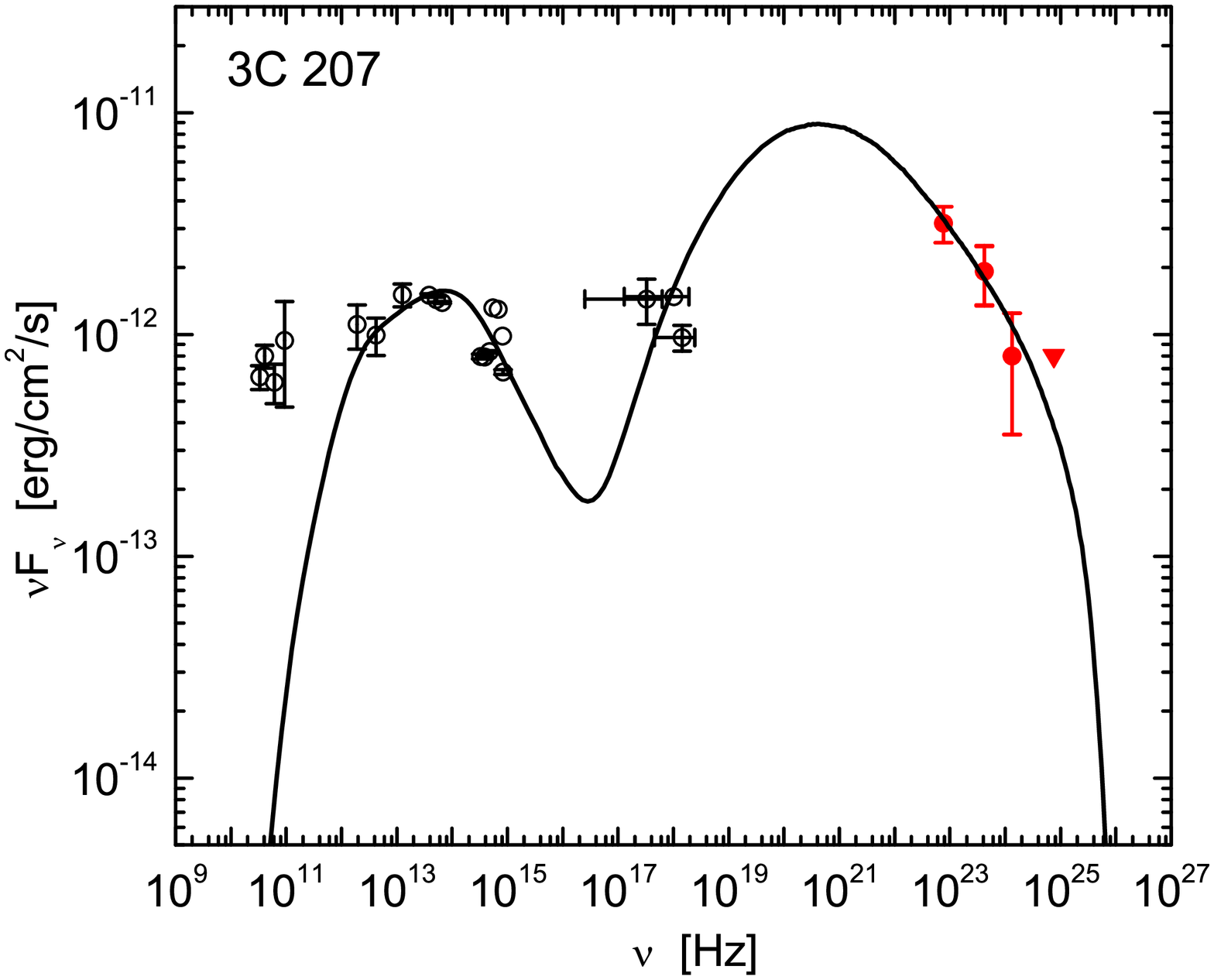}
\includegraphics[width=37mm,height=34mm]{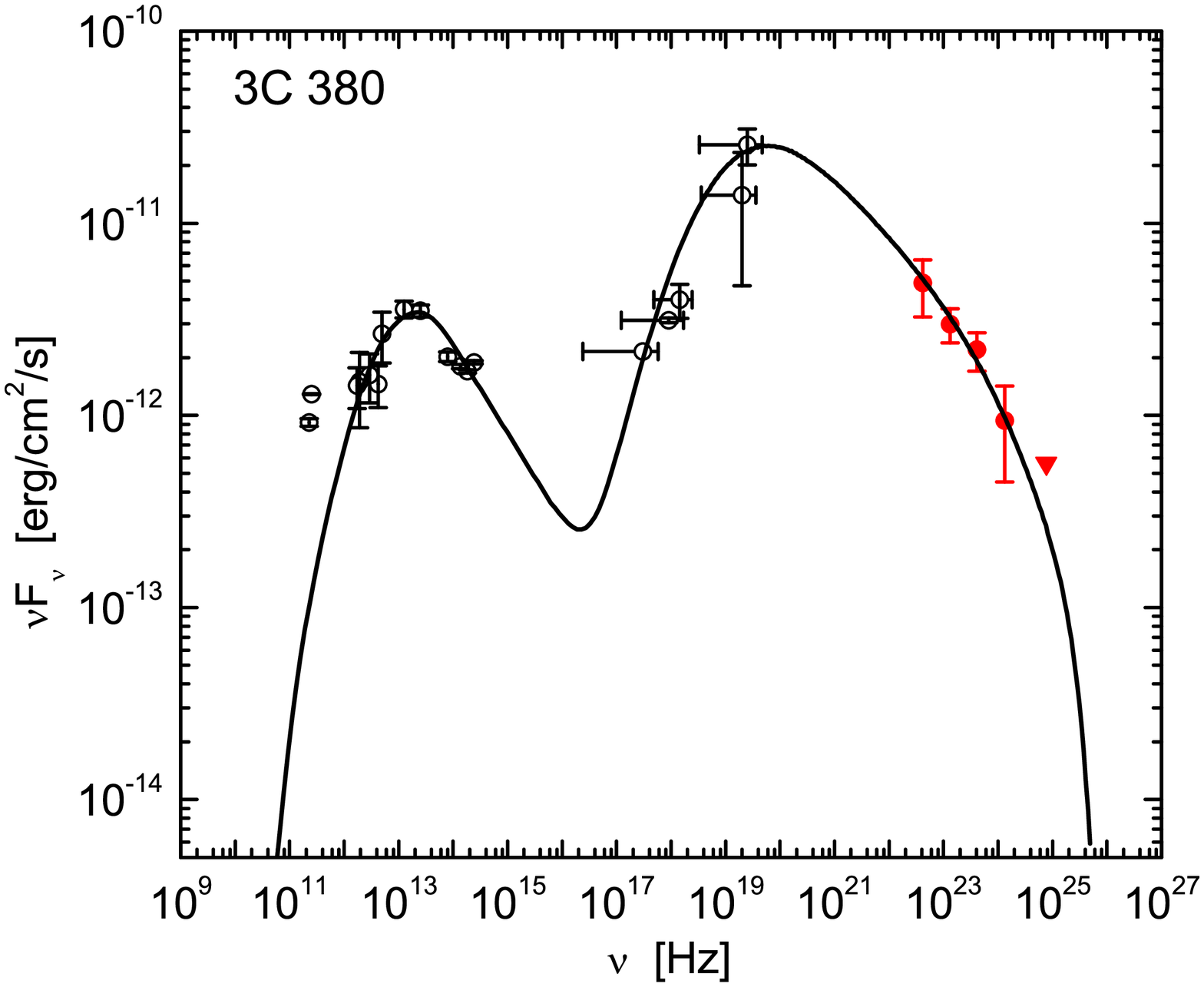}
\includegraphics[width=37mm,height=34mm]{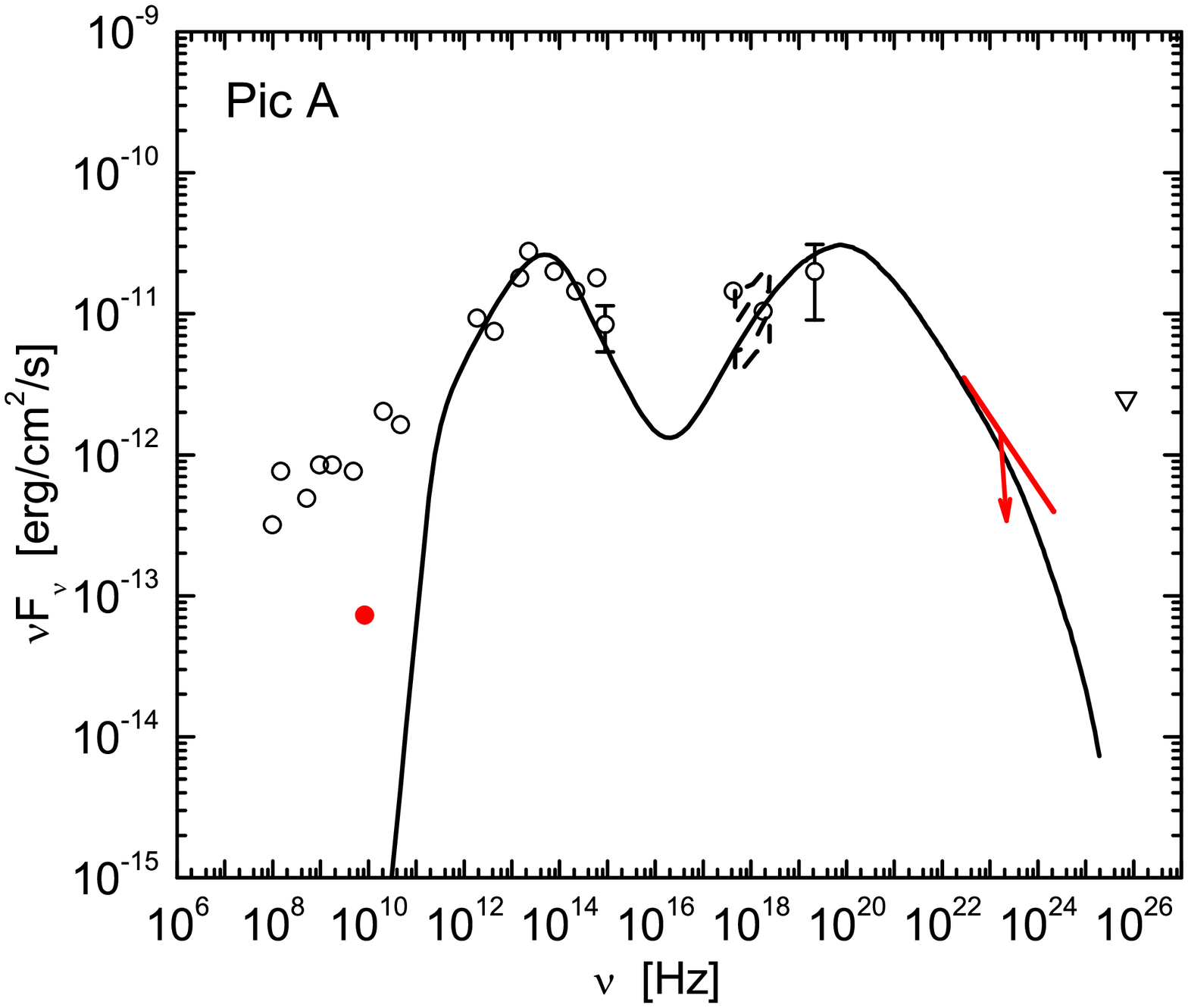}

\caption{The observed SEDs with the model fitting lines for the 12 RGs in our sample. The simultaneously or quasi-simultaneously observed data are presented as red solid symbols while the non-simultaneously observed data are marked as black open symbols or dashed lines. The triangles are upper-limits. The references for data on each of source are given in Table 1.}\label{SED}
\end{figure}

Fourteen RGs with confirmed redshift have been detected with \emph{Fermi}/LAT (Ackermann et al. 2015). The core radiation of Fornax A is very weak compared with their lobe, so the GeV emission of Fornax A may originate from the lobes (McKinley et al. 2015). For IC 310, there is no observation data available at lower-energy band, and the GeV--TeV spectra of this source cannot be represented with the simple one-zone leptonic model, as shown in the figure 5 in  Aleksi\'{c} et al. (2014). Therefore twelve GeV RGs with observed SEDs are included in our sample; eight FR I RGs (Cen A, NGC 1275/3C 84, M87, Cen B, PKS 0625--35, NGC 6251, NGC 1218, 3C 120) and four FR II RGs (3C 207, 3C 380, 3C 111, Pictor A). Their SEDs are collected and compiled from literature and NASA/IPAC Extragalactic Database (NED)\footnote{Note that the data taken from the NED for 3C 207 and 3C 380 should be the total emission of sources, including the emission from large-scale jets, which would result in the overestimation of the synchrotron radiation for the two sources.}, as shown in Figure~\ref{SED}; the simultaneously or quasi-simultaneously observed data are presented as red solid symbols while the non-simultaneously observed data are marked as black open symbols or dashed lines. The references for data on each of source are given in Table 1.

\begin{table}
\bc
\begin{minipage}[]{40mm}
\caption[]{SED Fitting Results\label{tab1}}\end{minipage}
\setlength{\tabcolsep}{4pt}
\small
 \begin{tabular}{ccccccccccccccccc}
  \hline\noalign{\smallskip}
Source &   $z$  &  $\delta$  &   $B$ &   $\Delta t$  &  $N_{0}$ &   $p_1$ &   $p_2$  &   $\gamma_{\min}$ &  $\gamma_{\rm b}$ &   $\gamma_{\max}$ &   $\theta$ &  Re &  $\log P_{\rm jet}$ &  $\log P_{\rm cav}$\\
   &    &   &  [G] &   [h] &   [cm$^{-1}$] &   &  &    &   &   [$\times\gamma_{\rm b}$] &  [deg] &   &  [erg/s] &  [erg/s]\\
  \hline\noalign{\smallskip}
NGC 1218&0.029&5.6&0.23&24&6.5E5&2.7&3.74&800&5.0E4&100&10.2&F15&42.99&43.79\\
NGC 1275&0.0179&5.8&0.15&168&1.4E1&1.4&3.2&250&1.3E3&200&9.9&A09a&44.31&43.34\\
NGC 6251&0.02471&7.8&0.02&24&5.3E6&2.7&4.3&300&1.6E4&10&7.3&M11&44.95&43.64\\
3C 120&0.033&1.8&3.7&120&8.0E6&2.76&4.76&260&1.9E3&50&31.8&K11&44.54\\
PKS 0625--35&0.055&4.9&1.2&24&9.5E1&1.74&3.5&1&2.0E4&100&11.7&F15&43.42&\\
M87&0.00428&3.0&0.1&48&2.1E5&2.42&4.3&180&1.0E4&100&19.1&A09b&43.58&43.79\\
Cen A&0.00183&1.2&4.1&24&6.1E4&1.56&4.38&100&9.1E2&100&47.7&A10b&43.36&43.37\\
Cen B&0.0129&4.8&0.1&24&1.9E5&2.2&3.7&1&3.0E3&100&11.9&K13&44.07&\\
3C 111&0.0485&4.7&0.45&24&2.8E4&1.7&4.8&300&2.0E3&100&12.2&K11&44.63&44.05\\
3C 207&0.681&9.8&0.42&24&7.9E6&2.56&4&300&3.3E3&100&5.8&A10a; NDE&45.53\\
3C 380&0.692&8.0&0.9&24&2.1E5&1.84&3.92&300&1.0E3&100&7.2&A10a; NDE&45.66&45.56\\
Pic A&0.0351&2.5&4.2&24&1.6E4&1.6&4.42&1&1.0E3&100&22.9&K11&43.88\\

  \noalign{\smallskip}\hline
\end{tabular}
\ec
\tablecomments{0.86\textwidth}{ $\theta$: Derived viewing angle.\\
Re: the references of SED data --- F15: Fukazawa et al. (2015); A09a: Abdo et al. (2009a); K11: Kataoka et al. (2011); A10a: Abdo et al. (2010a); NED: NASA/IPAC Extragalactic Database; A09b: Abdo et al. (2009b); A10b: Abdo et al. (2010b); M11: Migliori et al. (2011); K13: Katsuta et al. (2013).\\
$\log P_{\rm jet}$: For PKS 0625--35, Cen B, and Pic A, we take $\gamma_{\rm min}=300$ to calculate their jet powers, which is the median (the mean is $\gamma_{\rm min}=310$) of other nine RGs. \\
$\log P_{\rm cav}$: The cavity kinetic powers of RGs in Meyer et al. (2011).}
\end{table}

\section{SED Modeling and Results}

As shown in Figure \ref{SED}, the observed SEDs of RGs are similar to that of blazars and are double peaked. The observed SEDs of blazars can be explained with the one-zone leptonic model and this model is also used to reproduce the SEDs of RGs by many authors, e.g., synchrotron self-Compton (SSC) scattering process (Abdo et al. 2009a; Migliori et al. 2011; Fukazawa et al. 2015) and external inverse Compton (EC) scattering process (for M87, Cui et al. 2012). Although a more complex model with more parameters may present a better fit to the SEDs of some RGs (e.g., Tavecchio \& Ghisellini 2014), we prefer to use a simple model to fit the SEDs and to perform a statistical analysis of the jet properties on the basis of fitting results, and then compare their jet properties with that of blazars in our previous works (Zhang et al. 2012, 2014, 2015). Hence the one-zone leptonic model is used to fit the observed SEDs of these GeV RGs in this paper, where the model includes synchrotron radiation and SSC process. According to the unification models, the FR I and FR II RGs are parent populations of BL Lacs and FSRQs, respectively, and FR II RGs generally have stronger emission lines than FR I RGs, similar to FSRQs. The inverse Compton scattering of the photons from broad-line region (BLR, or torus) by the relativistic electrons in jets is wildly used to explain the gamma-ray emission of FSRQs. Different from the FSRQs, the radiations of FR II RGs in both X-ray and GeV bands can be well explained with one SSC component, hence we do not consider the EC/BLR process for FR II RGs during the SED fitting in this paper\footnote{We checked another case, taking the EC/BLR process into account during the SED fitting for the four FR II RGs. In this scenario, we will obtain a smaller $\delta$ value and a larger $B$ value, but a similar $\Gamma$ value for the four FR II RGs, and the following results in our paper still hold.  }.

There are nine parameters in this model. The radiation region is assumed as a homogenous sphere with radius $R$, magnetic field strength $B$, and Doppler factor $\delta$. The radius is obtained with $R = \delta c\Delta t/(1 + z)$, where $\Delta t$ is the variability timescale and listed in table 1, $c$ is the velocity of light, and $z$ is the redshift of each source. Similar to blazars, the GeV RGs have different variability timescales in different energy bands (e.g., Aleksi\'{c} et al. 2014). So we use the variability timescale at $\gamma$-ray band to constrain the emission region scale. For those RGs, which have no timescale available in literature, we take $\Delta t=1$ d. The electron distribution is taken as a broken power law, which is characterized by an electron density parameter ($N_{0}$), a break energy $\gamma_{\rm b}$, and indices ($p_1$ and $p_2$) in the range of $\gamma_{\rm e}$ to [$\gamma_{\min}$, $\gamma_{\max}$].

There are no enough simultaneous observation data to constrain the parameters as done for blazars in our previous works (Zhang et al. 2012, 2014, 2015), so the goodness of SED fitting is assessed visually. During the process of SED fitting, $p_1$ and $p_2$ are derived with the spectral indices of the observed SEDs as reported by Zhang et al. (2012). $\gamma_{\rm max}$ is fixed at 100$\gamma_{\rm b}$ and sometimes slightly varies to fit the SEDs. $\gamma_{\rm min}$ varies from the minimum value of unity until it can well explain the SEDs. The derived values of $\gamma_{\rm min}$, $\gamma_{\rm b}$, $N_{0}$, $B$, and $\delta$ are visually assessed, and may thus not be unique. The Klein--Nashina effects and the absorption of high energy gamma-ray photons by extragalactic background light (Franceschini et al. 2008) are also taken into account in our model calculations. The results of SED fitting are shown in Figure \ref{SED}, and the derived model parameters are reported in Table 1.

We find that six SEDs of RGs in our sample are also explained with the one-zone leptonic model by other authors (Fukazawa et al. 2015; Abdo et al. 2009a, 2009b, 2010; Migliori et al. 2011), and thus we compare the derived parameters with that reported in literature for these sources. Since the different radiation region sizes are taken for these SEDs, the derived parameter values are also slightly different. On the other hand, the model parameters are not independent, as discussed in Zhang et al. (2012). For example, $B$ and $\delta$ are dependent on each
other, i.e., $B\propto\delta^{-2}$ to $\delta^{-3}$, which is also consistent with the theoretical result in Tavecchio et al. (1998). The tighter
constraints on $B$ and $\delta$ would be obtained if the two peaks in the SEDs can be well constrained with the observation data (Zhang et al. 2012).

A simultaneous SED provides the constraint on the model parameters at a given state, and then provides a snapshot of the emitting population of particles at that time (Zhang et al. 2014, see also Bartoli et al. 2016). Note that the observation data from radio to $\gamma$-ray band of our sample sources are not totally simultaneous. The variation of flux at any energy band would result in the different values of model parameters (see also Fukazawa et al. 2016). Although a jet origin of the observed X-ray emission is suggested for some sources (Fukazawa et al. 2015), some RGs may also have the strong emission of accretion disk (Kataoka et al. 2007; Tanaka et al. 2015), and thus their optical and X-ray emission may not completely come from the jet radiation. However, we cannot avoid these effects with the limited observation data and also do not consider the effects in the following discussion.

\section{Comparisons of Jet Properties between GeV RGs and Blazars}

We have studied the jet properties of GeV blazars in our previous works (Zhang et al. 2012, 2014, 2015). It is thought that RGs are the parent populations of blazars (Urry \& Padovani 1995), hence we compare the jet physical properties between GeV RGs and blazars in this section, where the data of blazars are taken from Zhang et al. (2012, 2014, 2015). For BL Lacs, only the ones whose jet parameters can be constrained by SED fitting in Zhang et al. (2012) are considered and then there are 24 SEDs as cited in Zhang et al. (2014). For 30 FSRQs, we take the data in Zhang et al. (2015)\footnote{PKS 2142-758 in Zhang et al. (2015) is removed from our sample for the same reason as reported in Zhu et al. (2016).} for the same source in Zhang et al. (2014, 2015). Note that there are only twelve RGs in our sample, including eight FR I RGs and four FR II RGs, so the sample of RGs is very limited.

\subsection{Jet Parameters}

\begin{figure}\centering
\includegraphics[angle=0,scale=0.3]{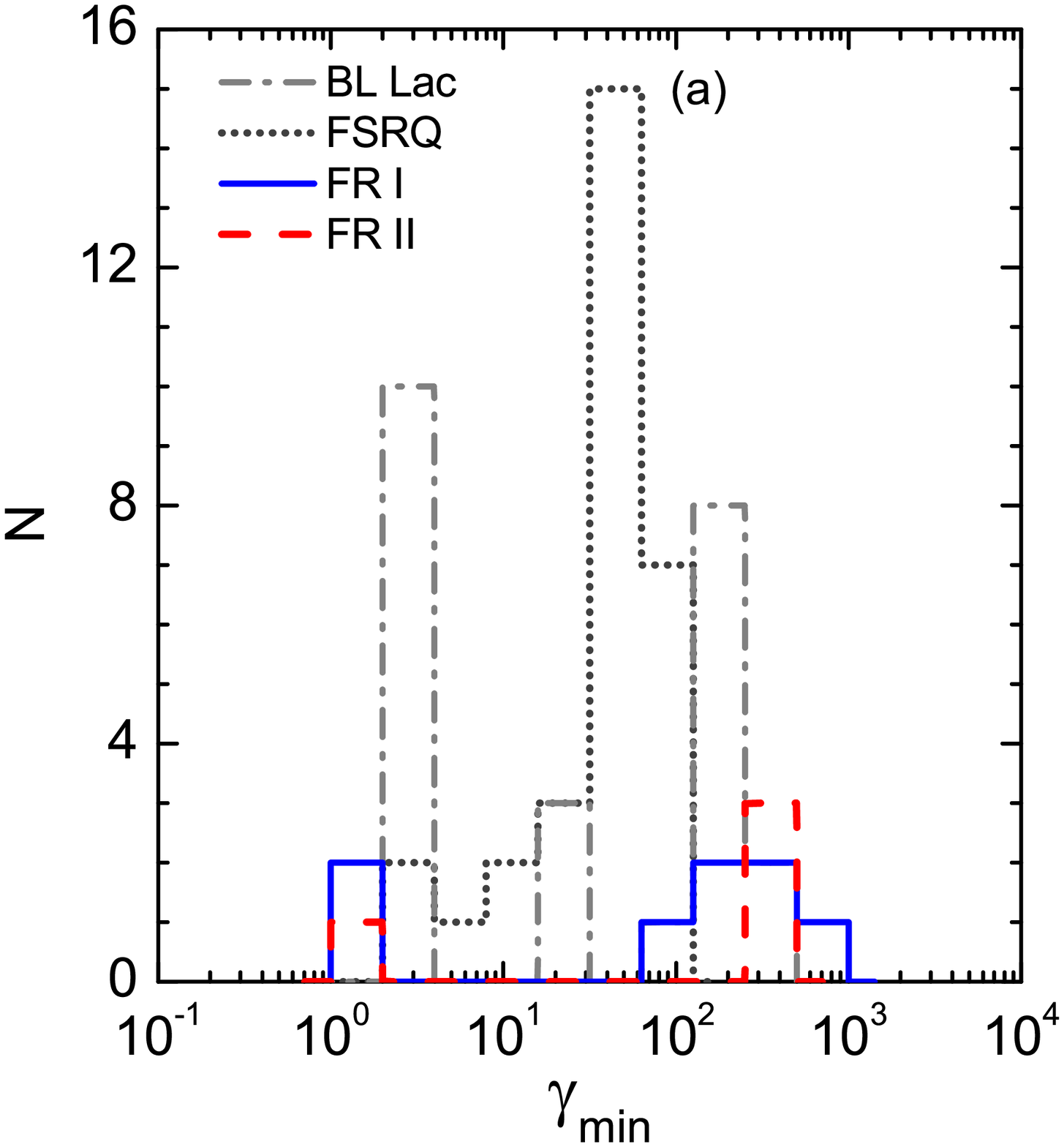}
\includegraphics[angle=0,scale=0.3]{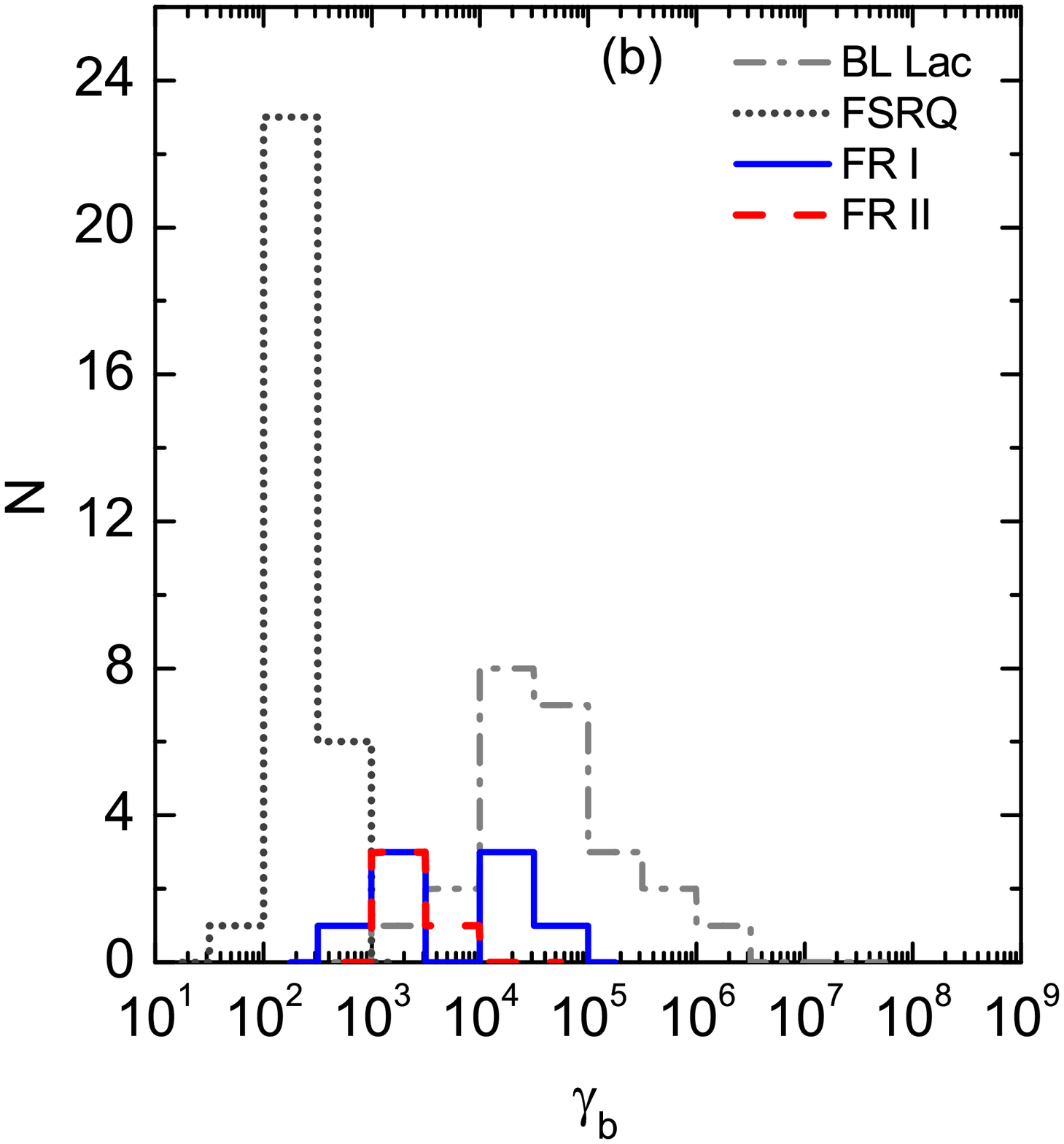}\\
\includegraphics[angle=0,scale=0.3]{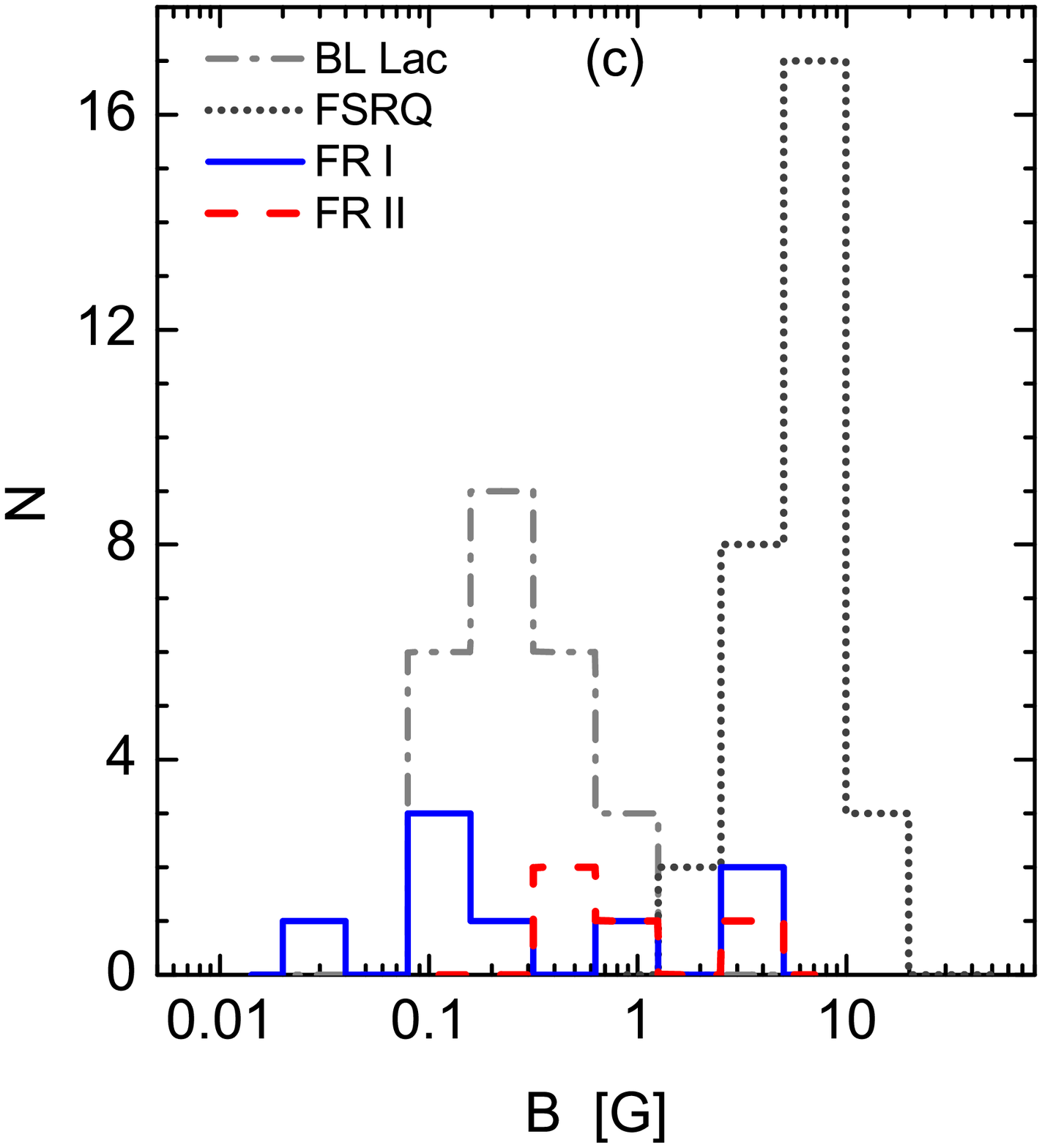}
\includegraphics[angle=0,scale=0.3]{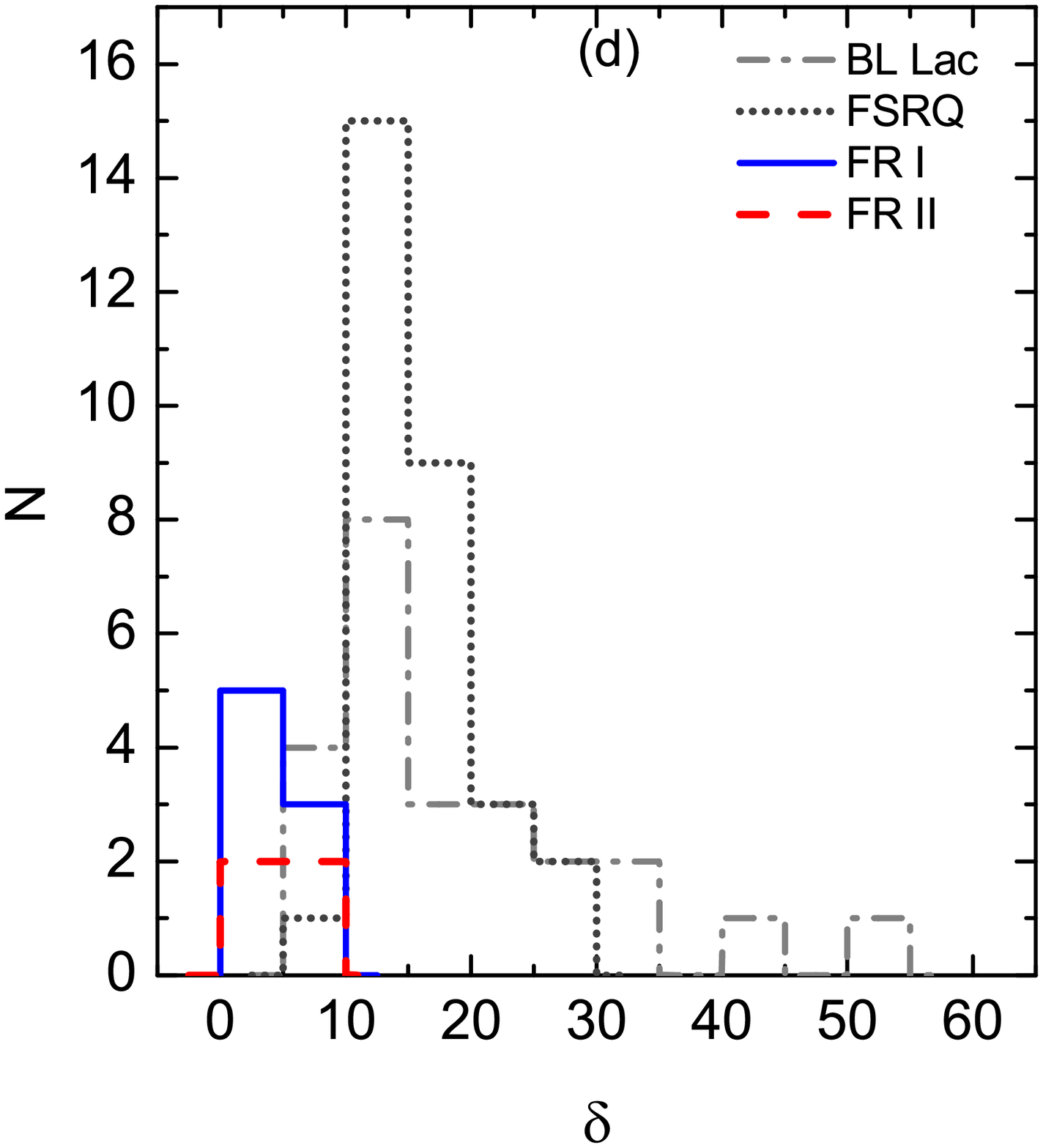}
\caption{Distributions of the minimum Lorentz factor of electrons ($\gamma_{\rm min}$, \emph{Panel a}), the break Lorentz factor of electrons ($\gamma_{\rm b}$, \emph{Panel b}), the magnetic field strength ($B$, \emph{Panel c}), and the beaming factor ($\delta$, \emph{Panel d}). The blazar data are taken from Zhang et al. (2012, 2014, 2015).}\label{Dis-para}
\end{figure}

The distributions of the derived jet parameters by SED fitting are shown in Figure \ref{Dis-para}. For most of the GeV RGs, the minimum energies of electrons ($\gamma_{\min}$) in jets are higher than unity as given in Figure \ref{Dis-para}(a), similar to blazars (Zhang et al. 2012, 2014, 2015). The $\gamma_{\rm b}$ distributions for both FR I and FR II RGs roughly cover the intermediate region of BL Lacs and FSRQs, but on average FR I RGs have larger $\gamma_{\rm b}$ than FR II RGs, as given in Figure \ref{Dis-para}(b), which is also observed between FSRQs and BL Lacs. The magnetic field strength of RGs is lower than that of FSRQs and more similar to that of BL Lacs, but on average FR II RGs have higher $B$ than FR I RGs, as presented in Figure \ref{Dis-para}(c). The $\delta$ values for most of RGs are lower than that of blazars as shown in Figure \ref{Dis-para}(d). This is consistent with the unification model of RL AGNs that RGs are the misaligned parent populations of blazars with smaller Doppler factors. However, the $\delta$ values of some RGs are similar to that of blazars with $\delta\sim10$, and these RGs with larger Doppler factors may also have small viewing angles of jets like blazars. Hence maybe only a fraction of RGs that are thought to be the parent population of blazars have been detected in VHE $\gamma$-ray band (Aharonian et al. 2006, 2009; Aleksic et al. 2012; Tavecchio \& Ghisellini 2014).

Note that if the viewing angle is larger than the opening angle of the jet, the one-zone leptonic model would not be able to explain the observation
data of blazars, as discussed in Zhang et al. (2015). Hence, the viewing angle should be smaller than the opening angle of the jet in blazars, and the probability is highest when looking at the jet at the angle of $1/\Gamma$, where $\Gamma$ is the bulk Lorenz factor. When the viewing angle ($\theta$) is equal to the opening angle ($1/\Gamma$) of a jet, we obtain $\Gamma=\delta$ and then we can derive the values of viewing angle\footnote{For RGs, it should be the lower limit of viewing angle.} with the $\delta$ values. The derived values of viewing angle are also given in Table 1. $\theta$ ranges from $\sim5^{\circ}$ to $\sim48^{\circ}$ with a mean of $\sim16^{\circ}$. With the derived Doppler factors by flux density variation at radio band and the apparent jet speed, Hovatta et al. (2009) calculated the viewing angles of six RGs (III ZW2, 3C 84, 3C 111, 3C 120, 3C 380, OW 637), which range from $\sim7^{\circ}$ to $\sim40^{\circ}$ with a mean of $\sim20^{\circ}$. This is roughly consistent with our results. Hovatta et al. (2009) also reported that the derived Doppler factors of RGs are much smaller than that of blazars, but the viewing angles are larger than that of blazars.

\subsection{Jet Power and Cavity Kinetic Power}

Based on the jet parameters of SED fitting and the assumption of $\Gamma=\delta$, we also calculate the jet power ($P_{\rm jet}$). It is assumed that the jet power of RGs is carried by relativistic electrons, cold protons, magnetic fields, and radiations, i.e., $P_{\rm jet}=\sum_i\pi R^2 \Gamma^2 c U^{'}_{\rm i}$, where $U^{'}_{i} (i={\rm e,\ p,}\ B, \rm r)$ are the energy densities associated with the emitting electrons
($U^{'}_{\rm e}$), cold protons ($U^{'}_{\rm p}$), magnetic fields ($U^{'}_{B}$), and radiations ($U^{'}_{\rm r}$) measured in the co-moving frame (Ghisellini et al. 2009).
Following our and other author's works about blazars (e.g., Ghisellini et al. 2009; Zhang et al. 2012, 2014), the proton-electron pair assumption is also used here. The values of jet powers are also reported in Table 1. Note that the $\gamma_{\rm min}$ values of PKS 0625--35, Cen B, and Pic A are taken as $\gamma_{\rm min}=1$ in Table 1, which may overestimate the values of $P_{\rm jet}$. Hence we take $\gamma_{\rm min}=300$ to calculate their jet powers and the powers of electrons and protons, where $\gamma_{\rm min}=300$ is the median of the other nine GeV RGs.

\begin{figure}\centering
\includegraphics[angle=0,scale=0.35]{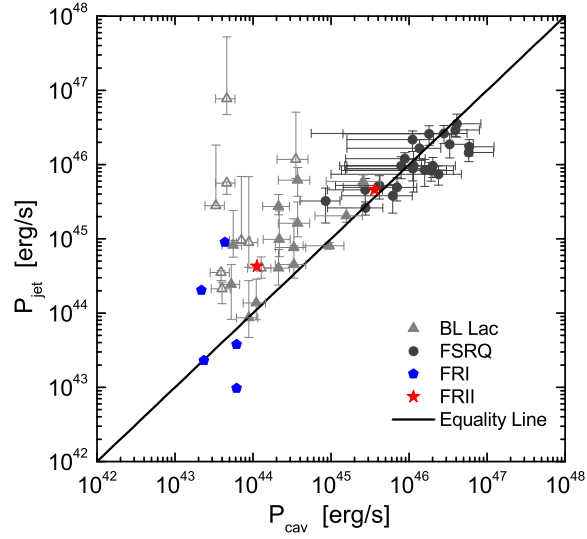}
\caption{Comparison between $P_{\rm jet}$ with $P_{\rm cav}$. The blue pentagons and red stars are for FR I and FR II RGs, respectively. The data of FSRQs (dark-gray circles) and BL Lacs (gray triangles) from Zhang et al. (2014) are also presented, where the opened gray triangles are for BL Lacs with $\gamma_{\min}=2$ as reported in Zhang et al. 2014. The solid line is the equality line.}\label{Lkin-Pjet}
\end{figure}

The RGs always have large-scale jets, which are believed to be connected with their central engines (Harris \& Krawczynski 2006). The observed X-ray cavities are evidence for AGN feedback and provide a direct measurement of the mechanical energy released by AGNs (B\^{\i}rzan et al. 2008; Cavagnolo et al. 2010). The cavity kinetic power is correlated with the radio power in large-scale of galaxies (B\^{\i}rzan et al. 2004; 2008; Cavagnolo et al. 2010; O'Sullivan et al. 2011), and thus the cavity kinetic power ($P_{\rm cav}$) can be estimated using the relation between $P_{\rm cav}$ and radio luminosity. There are five FR I RGs (NGC 1218, NGC 1275, NGC 6251, M 87, Cen A) and two FR II RGs (3C 111 and 3C 380) in our sample with available $P_{\rm cav}$ in Meyer et al. (2011), as listed in Table 1. Comparison between $P_{\rm jet}$ and $P_{\rm cav}$ for the GeV RGs is given in Figure \ref{Lkin-Pjet}, and the data of FSRQs and BL Lacs from Zhang et al. (2014) are also presented. The distributions of RGs in the $P_{\rm cav}$--$P_{\rm jet}$ plane are roughly consistent with the blazars; the five FR I RGs are in the low power end of BL Lacs, while one FR II RG overlaps with the distributions of FSRQs and another one overlaps with BL Lacs. However, with the small sample we cannot suggest that the FR I RGs are unified with BL Lacs and FR II RGs are unified with FSRQs in the $P_{\rm cav}$--$P_{\rm jet}$ plane.

\subsection{Jet Composition and Radiation Efficiency}

\begin{figure}
\includegraphics[angle=0,scale=0.23]{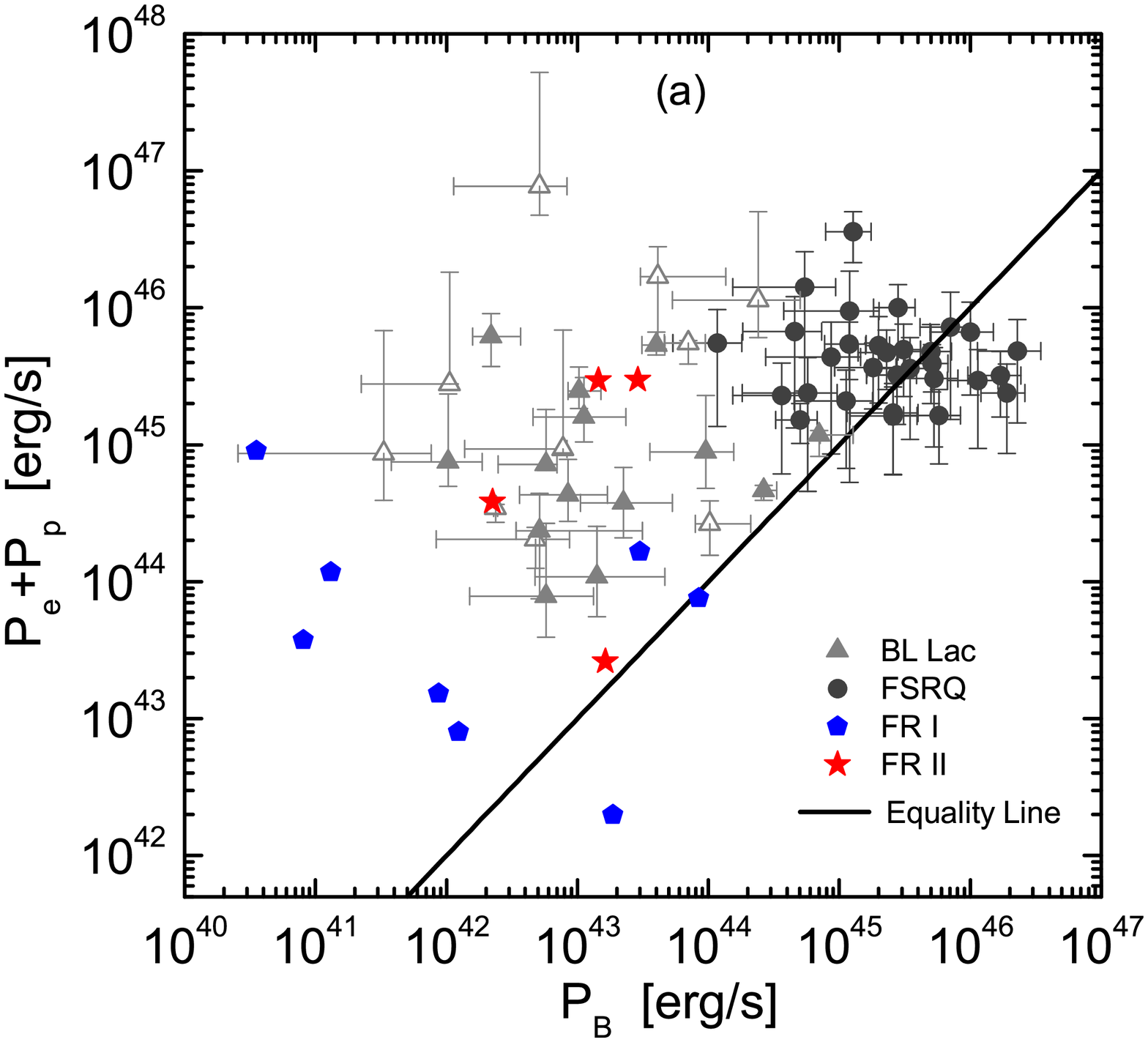}
\includegraphics[angle=0,scale=0.23]{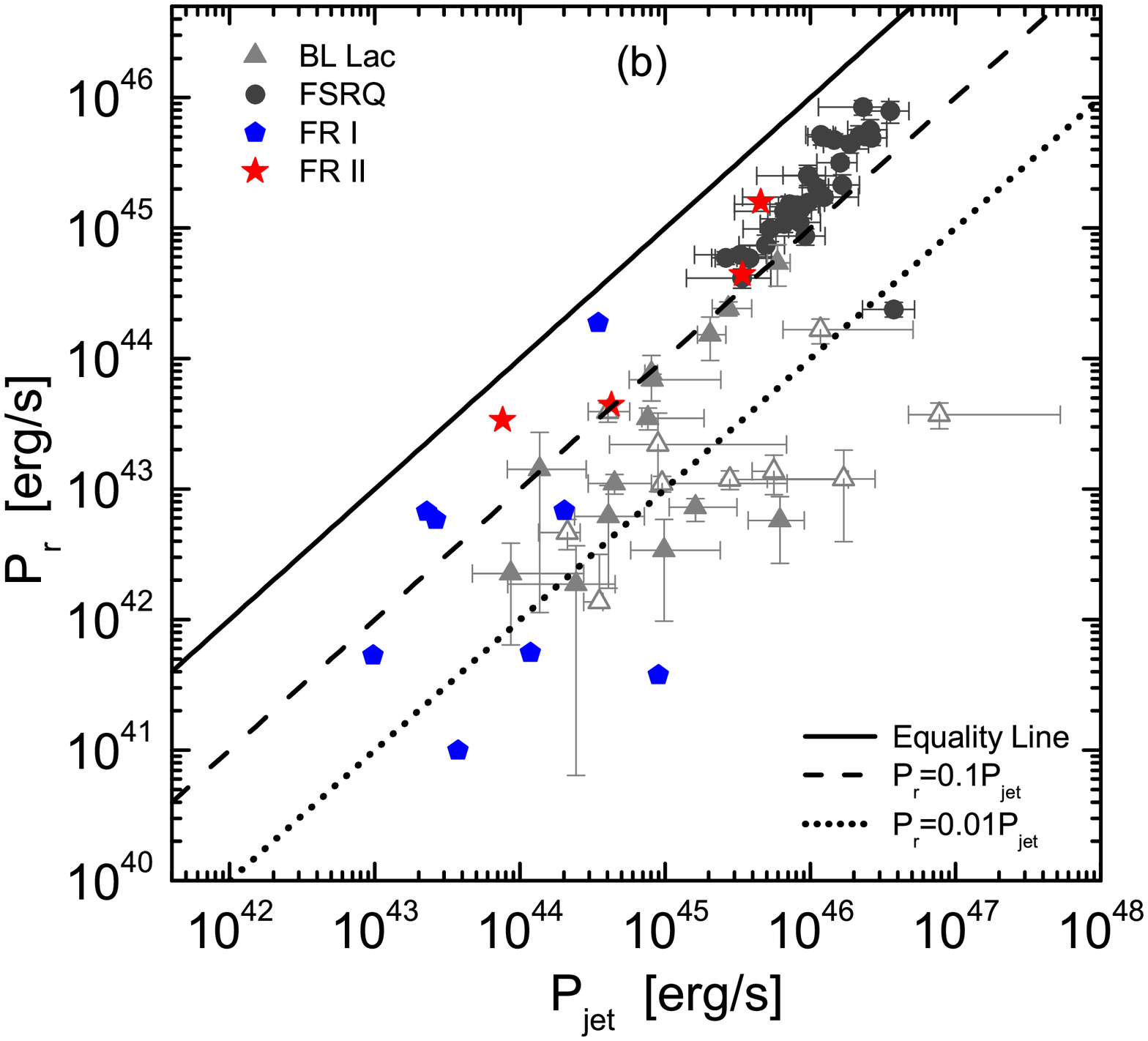}
\includegraphics[angle=0,scale=0.23]{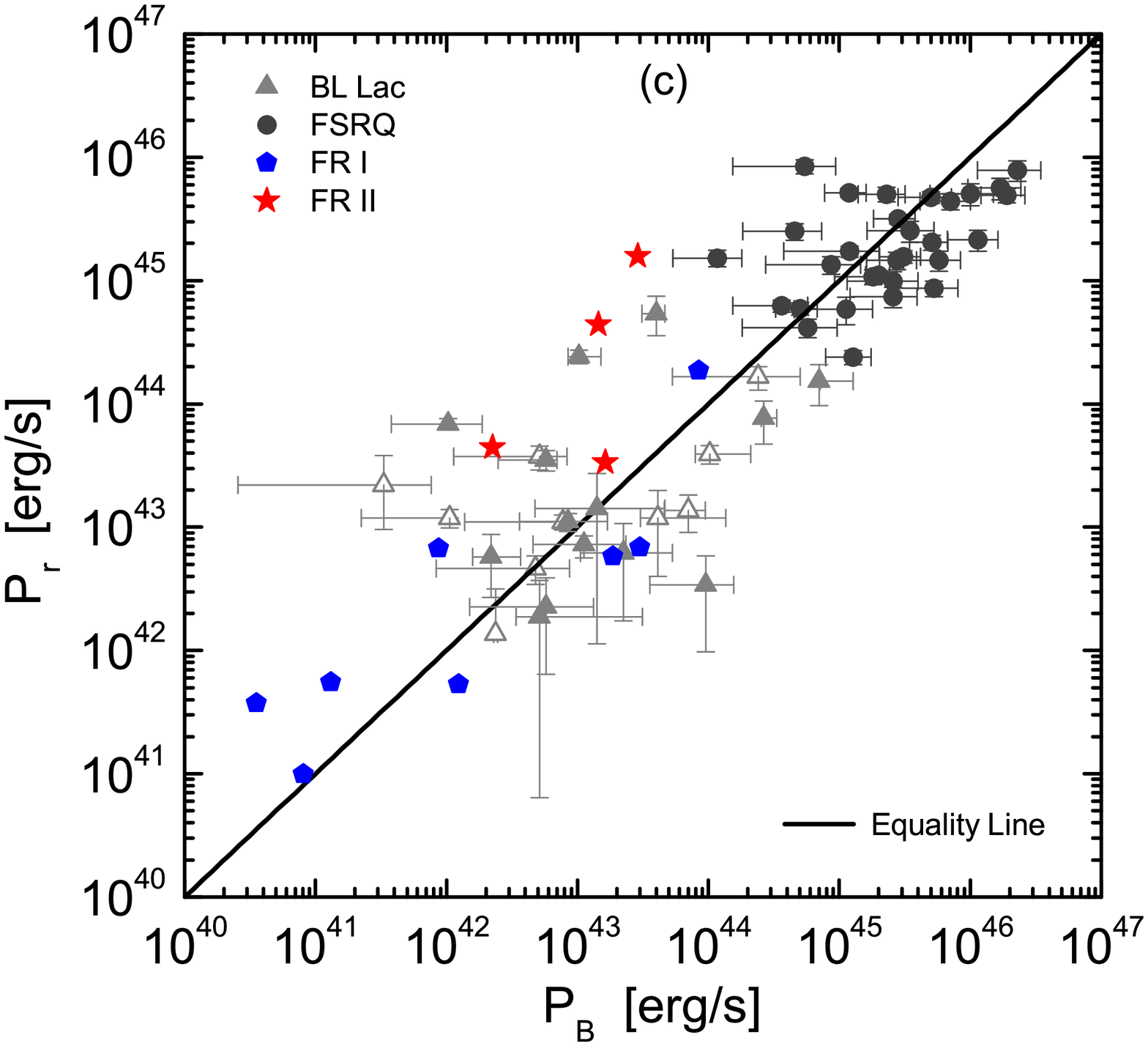}
\caption{$P_{\rm e}+P_{\rm p}$ as a function of $P_{B}$, and $P_{\rm r}$ as functions of $P_{\rm jet}$ and $P_{B}$. The dark-gray circles are for FSRQs from Zhang et al. (2014, 2015). The gray triangles are for BL Lacs from Zhang et al. (2012) while the opened gray triangles are for BL Lacs with $\gamma_{\min}=2$ as reported in Zhang et al. 2014. The blue pentagons and red stars are for FR I and FR II RGs, respectively.}\label{Pjet}
\end{figure}

In order to investigate the jet composition and radiation efficiency of RGs, we also calculate the powers carried by each components; the powers of electrons ($P_{\rm e}$), protons ($P_{\rm p}$), magnetic fields ($P_{B}$), and radiations ($P_{\rm r}$). $P_{\rm e}+P_{\rm p}$ as a function of $P_{B}$, and $P_{\rm r}$ as the functions of $P_{\rm jet}$ and $P_{B}$ are shown in Figure \ref{Pjet}. The data of BL Lacs in Zhang et al. (2012) and FSRQs in Zhang et al. (2014, 2015) are also presented in Figure \ref{Pjet}. Only FR I RG PKS 0625--35 has much higher $P_{B}$ than its $P_{\rm e}+P_{\rm p}$ as shown in Figure \ref{Pjet}(a). Therefore, the jets of RGs are likely dominated by particles. This is different from the results of blazars: the FSRQ jets are highly magnetized and the BL Lac jets are matter dominated (Zhang et al. 2014). In the $P_{\rm jet}$--$P_{\rm r}$ plane, $P_{\rm r}$ for all the RGs is lower than their $P_{\rm jet}$, similar to blazars. The jet radiation efficiency $\epsilon_{\rm r}=P_{\rm r}/P_{\rm jet}$ for most of the GeV RGs is larger than 0.01, and the four FR II RGs have $\epsilon_{\rm r}>0.1$, which is similar to FSRQs. On average, $\epsilon_{\rm r}$ of FR II RGs is higher than that of FR I RGs. One can also observe that $\epsilon_{\rm r}$ for most of FSRQs ranges from 0.1 to 1 while $\epsilon_{\rm r}$ for most of BL Lacs is between 0.01 and 0.1. In this respect, it seems that FR I RGs are unified with BL Lacs with low jet radiation efficiency while FR II RGs are unified with FSRQs with high jet radiation efficiency. In the $P_{B}$--$P_{\rm r}$ plane, RGs roughly follow the distributions of blazars along the equality line and extend to the low power end. These results may indicate a possible correlation between $P_{B}$ and $P_{\rm r}$, suggesting that the radiation efficiency of jet may be related with jet magnetization for the GeV RGs, analogous to blazars.

\subsection{The Sequence in the $\nu_{\rm s}-P_{\rm jet}$ Plane}

\begin{figure}\centering
\includegraphics[angle=0,scale=0.3]{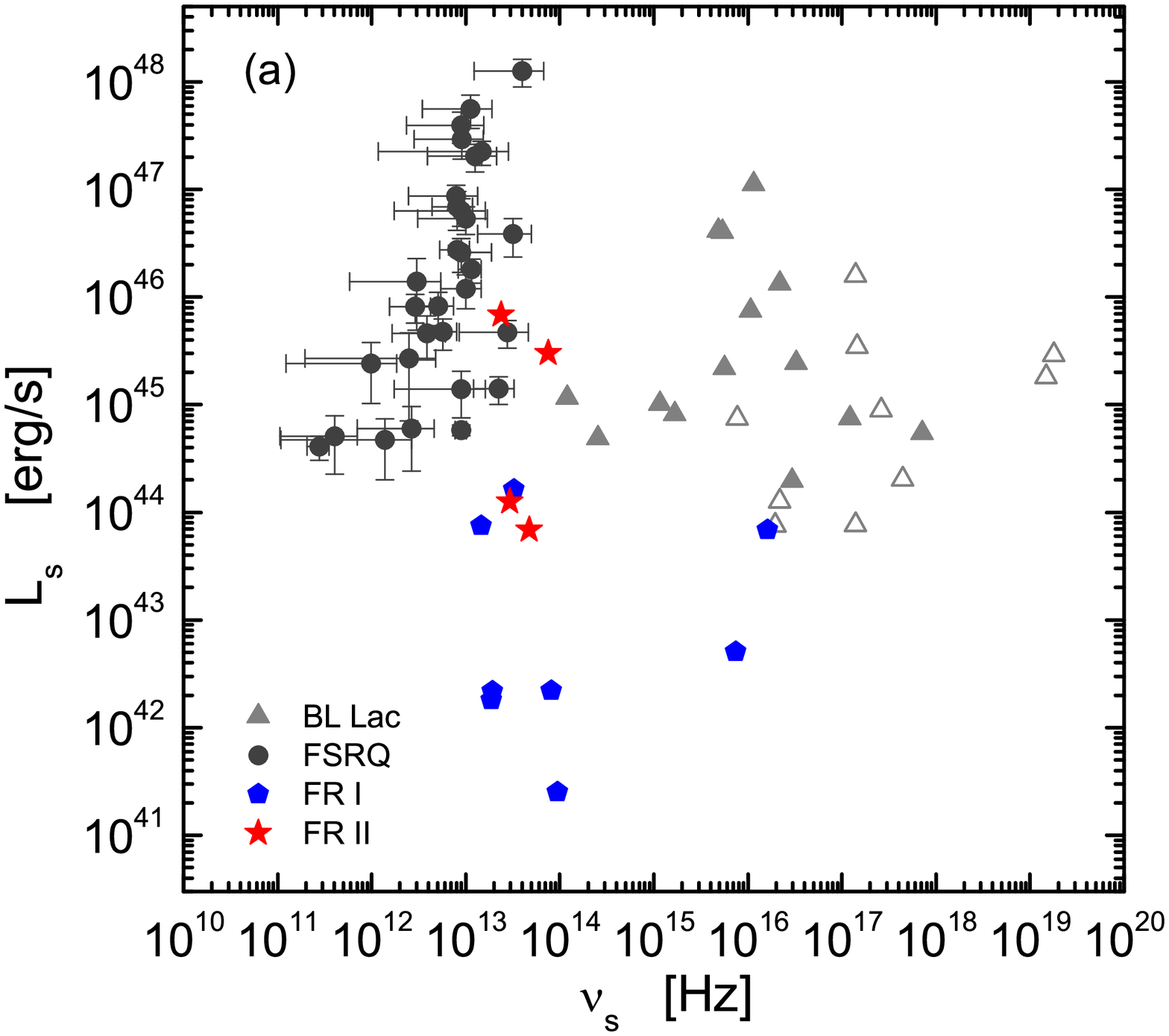}
\includegraphics[angle=0,scale=0.3]{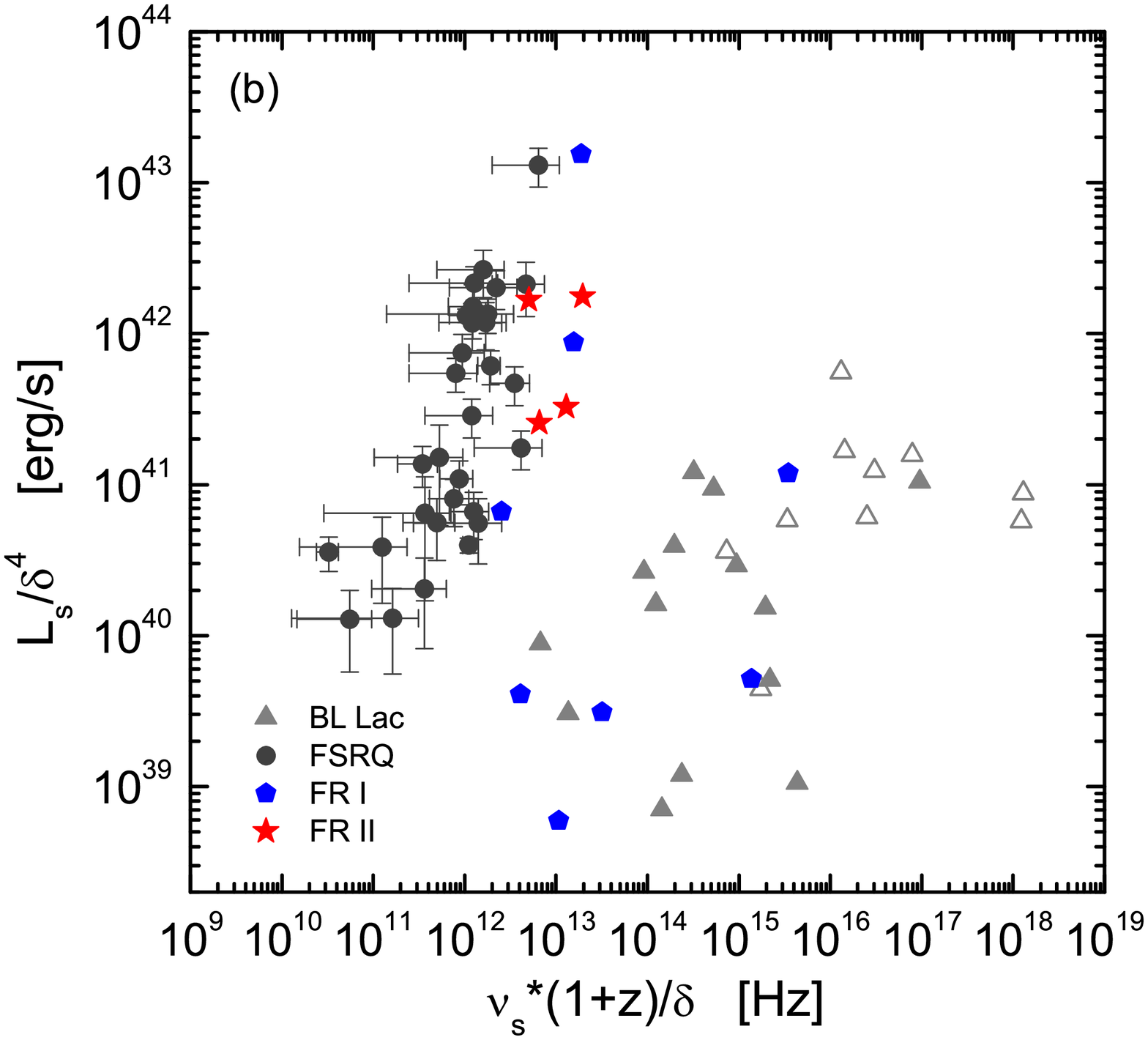}
\includegraphics[angle=0,scale=0.3]{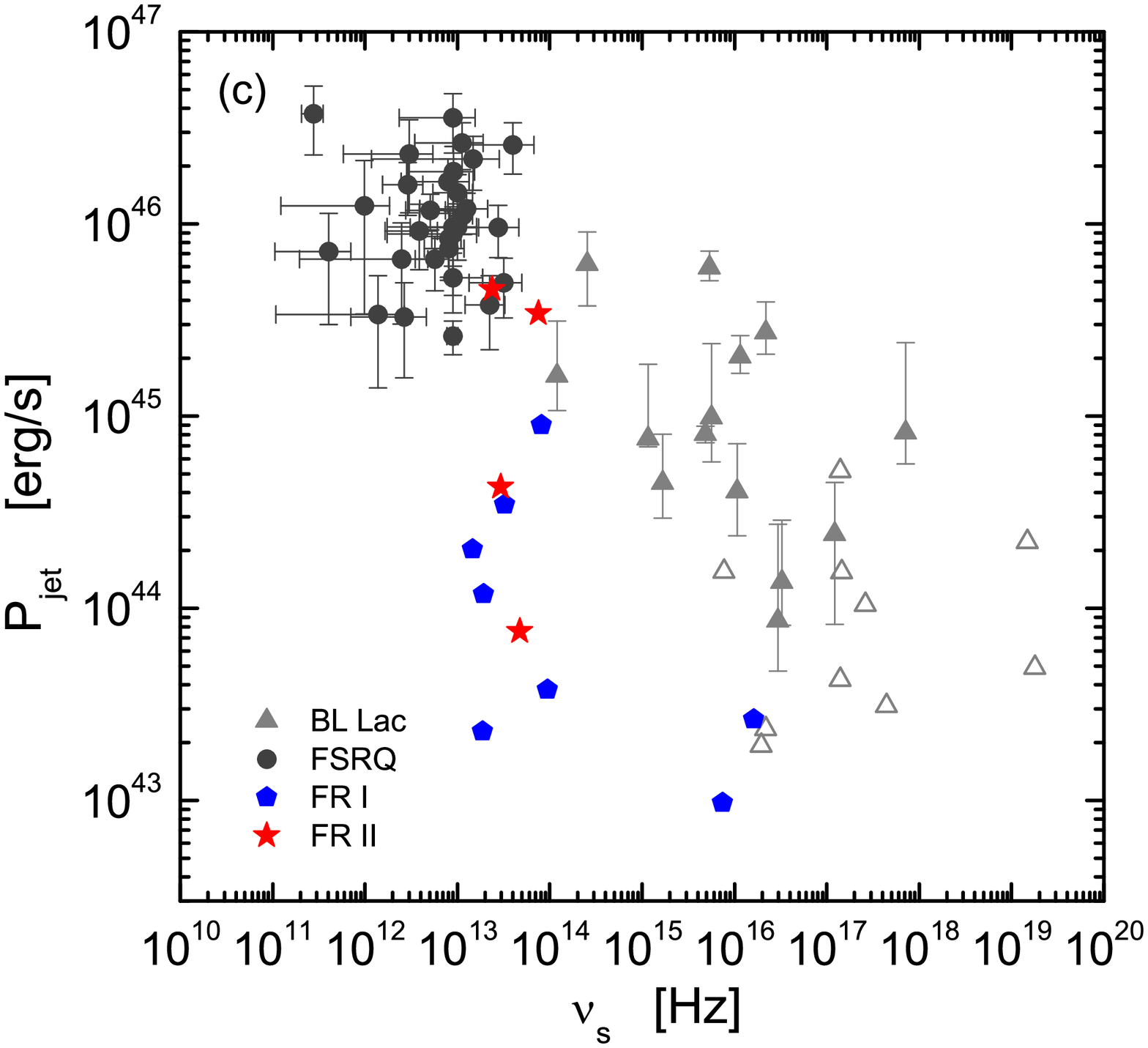}
\includegraphics[angle=0,scale=0.3]{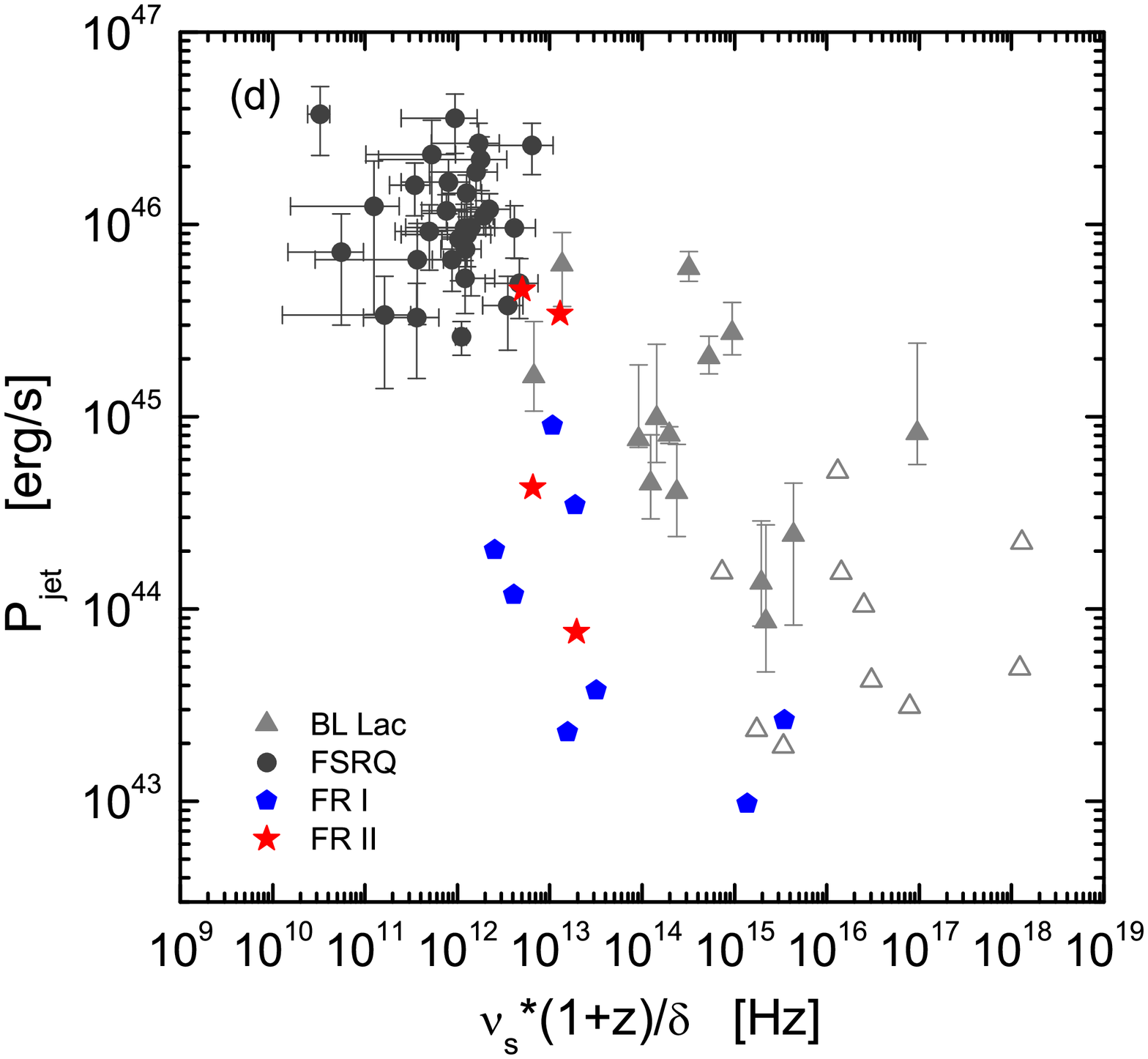}
\caption{Synchrotron peak luminosity ($L_{\rm s}$) and jet power ($P_{\rm jet}$) as a function of the synchrotron peak frequency ($\nu_{\rm s}$) in the observer (\emph{Panels (a), (c)}) and co-moving (\emph{Panels (b), (d)}) frames. Note that for BL Lacs with $\gamma_{\min}=2$ as reported in Zhang et al. 2014, their $P_{\rm jet}$ values are recalculated with $\gamma_{\min}=190$ and are marked as opened triangles, more details to see Section 4.4.}\label{sequence}
\end{figure}

The 12 GeV RGs and the GeV blazars in our sample are shown in the $\nu_{\rm s}-L_{\rm s}$ plane in Figure \ref{sequence}(a), where $\nu_{\rm s}$ and $L_{\rm s}$ are the peak frequency and peak luminosity of their synchrotron radiation, respectively. No significant evidence is found for the observation of the ``blazar sequence" \footnote{The increase of $\nu_{\rm s}$ corresponds to the decreases of the luminosity (Fossati et al. 1998).} with or without including the 12 GeV RGs. FSRQs and BL Lacs are clearly separated by $\nu_{\rm s}=10^{14}$ Hz. The distribution of the four FR II RGs is marginally consistent with FSRQs, while the eight FR I RGs have more similar $\nu_{\rm s}$ with BL Lacs, but much lower luminosity than BL Lacs. Considering the different Doppler factors of these sources, we show the $\nu_{\rm s}-L_{\rm s}$ relation in the co-moving frame in Figure \ref{sequence}(b). Still no track of ``blazar sequence" is found with or without including the 12 GeV RGs. However, the phenomenon of ``balzar envelope" suggested by Meyer et al. (2011) is observed for these GeV AGNs.

It becomes interesting if we replace $L_{\rm s}$ with the jet power ($P_{\rm jet}$) of sources, as given in Figures \ref{sequence}(c),(d). Note that the $\gamma_{\rm min}$ values for ten SEDs of BL Lacs are poorly constrained and taken as $\gamma_{\rm min}=2$ in Zhang et al. (2012), which may lead to significant overestimation of their $P_{\rm jet}$ as described in Section 4.2. So we use the median of $\gamma_{\rm min}=190$ for the other 14 SEDs of BL Lacs to recalculate the $P_{\rm jet}$ values for the ten SEDs of BL Lacs, which are shown as opened triangles without errors in Figures \ref{sequence}(c),(d). A strong anticorrelation between $\nu_{\rm s}$ and $P_{\rm jet}$ is found for all the data points in the plane with a Pearson correlation coefficient $r=-0.69$ and a chance probability $p=2.1\times10^{-10}$. After correcting the peak frequency into the co-moving frame, this correlation becomes even stronger as given in Figure \ref{sequence}(d) with a Pearson correlation coefficient $r=-0.74$ and a chance probability $p=2.1\times10^{-12}$. The distributions of the GeV RGs are roughly consistent with the distributions of these GeV blazars with slightly lower powers\footnote{Although the lower $P_{\rm cav}$ of RGs than that of blazars in Figure 3 may imply that RGs should have the lower jet powers, we cannot report that the jet powers of RGs should be intrinsic lower than that of blazars with the limited sample sources. And the kinetic energy of X-ray cavities only provides a lower limit to the jet energy if shocks exist in the hot gas (B\^{\i}rzan et al. 2008). Hence, we propose that the lower jet powers of RGs should be affected by the poor constraints on model parameters.}. These results indicate that the ``sequence" behavior among blazars, together with the GeV RGs, may be dominated by the jet power intrinsically. It means that the jet power regulates the synchrotron peak, as reported by Meyer et al. (2011).

\section{Summary}

A SED sample of 12 GeV RGs is collected and compiled from the literature and NED. On the basis of the derived jet parameters by SED fits with the one-zone leptonic model, we calculated the jet powers and the powers carried by each component to investigate their jet compositions and radiation efficiencies, as well as the relations between jet power and lager-scale kinetic power. We also presented a comparison of jet properties between the GeV RGs and blazars, where the data of blazars are taken from Zhang et al. (2012, 2014, 2015). Our results are summarized below.
\begin{itemize}

\item The observed SEDs of the 12 GeV RGs can be explained with the one-zone leptonic model, i.e., synchrotron + SSC model.

\item Their distributions of $B$, $\gamma_{\rm b}$, and $\gamma_{\min}$, span the parameter spaces of BL Lacs and FSRQs. No significant unification is found for these jet parameters between FR I RGs and BL Lacs and between FR II RGs and FSRQs. However, on average FR I RGs have the larger $\gamma_{\rm b}$ and lower $B$ than FR II RGs, analogous to the differences between BL Lacs and FSRQs. The derived $\delta$ of RGs is on average smaller than that of balzars, which is consistent with the unification model in which RGs are the misaligned parent populations of blazars with smaller Doppler factors.

\item In the $P_{\rm cav}$--$P_{\rm jet}$ plane, the distributions of RGs are roughly consistent with blazars, and extend to the low power end.

\item Most of the RG jets may be dominated by particles, but their jet radiation efficiencies could be still related with the extent of their jet magnetization. On average the jet radiation efficiencies of FR II RGs are higher than that of FR I RGs.

\item A strong anticorrelation between $\nu_{\rm s}$ and $P_{\rm jet}$ is observed for the GeV blazars and GeV RGs, and this correlation becomes stronger after correcting the peak frequency into the co-moving frame, indicating that the ``sequence" behavior among blazars, together with the GeV RGs, may be dominated by the jet power intrinsically.

\end{itemize}

\begin{acknowledgements}
We thank the anonymous referee for his/her valuable suggestions. We appreciate helpful discussion with Ting-Feng Yi. This work is supported by the National Natural Science Foundation of China (grants 11573034, 11533003, 11373036, 11133002), the National Basic Research Program (973 Programme) of China (grant 2014CB845800), and the Guangxi Science Foundation (2013GXNSFFA019001).
\end{acknowledgements}

\label{lastpage}

\end{document}